\documentclass[letterpaper,12pt]{article}

\usepackage[dvips]{graphicx}
\usepackage{ifthen}
\usepackage{enumerate}
\usepackage{cite}
\usepackage{amssymb}
\usepackage[mathscr]{eucal}
\usepackage[errorshow]{tracefnt}
\usepackage{amsmath}
\usepackage{amssymb}
\usepackage{amsthm}
\usepackage{eucal}
\usepackage{eufrak}
\usepackage{epsfig,multicol,bbm}
\usepackage{longtable}
\usepackage{amsthm,latexsym,multibox,amssymb,amsfonts,array}

\newcommand{\beq}{\begin{equation}}
\newcommand{\beqn}{\begin{equation*}}
\newcommand{\eeq}{\end{equation}}
\newcommand{\eeqn}{\end{equation*}}
\newcommand{\beqa}{\begin{eqnarray}}
\newcommand{\beqan}{\begin{eqnarray*}}
\newcommand{\eeqa}{\end{eqnarray}}
\newcommand{\eeqan}{\end{eqnarray*}}
\newcommand{\bdm}{\begin{displaymath}}
\newcommand{\edm}{\end{displaymath}}

\newcommand{\la}{\langle}
\newcommand{\ra}{\rangle}

\newcommand{\ba}{\begin{array}}
\newcommand{\ea}{\end{array}}

\newcommand\ffam{\sffamily}
\newcommand\fser{\bfseries}

\newcommand\nn{\nonumber}

\newcommand\benu{\begin{enumerate}}
\newcommand\eenu{\end{enumerate}}
\newcommand\bit{\begin{itemize}}
\newcommand\eit{\end{itemize}}

\newtheorem{hej}{Theorem}[section]

\def\Pf{\noindent \textbf{Proof. }}

\def\End{\mathrm{End\,}}

\def\tr{\mathrm{tr\,}}

\def\dim{\mathrm{dim\,}}

\def\der'{\mathfrak{der}'\,}
\def\der{\mathfrak{der}\,}
\def\str'{\mathfrak{str}'\,}
\def\str{\mathfrak{str}\,}
\def\con{\mathfrak{con}\,}
\def\R{\mathbb{R}}

\def\C{\mathbb{C}}

\def\H{\mathbb{H}}

\def\bbO{\mathbb{O}}
\def\upi{\textup{i}}

\def\frake{\mathfrak{e}}
\def\f{\mathfrak{f}}
\def\g{\mathfrak{g}}
\def\h{\mathfrak{h}}
\def\so{\mathfrak{so}}

\def\e,n{e_1,\,e_2,\,\ldots,\,e_n}

\def\a,n{a_1,\,a_2,\,\ldots,\,a_n}

\def\al,n{\alpha_1,\,\alpha_2,\,\ldots,\,\alpha_n}

\def\qed{\hspace{\stretch{1}} $\square$ \\

\noindent}

\RequirePackage{hyperref}

\numberwithin{equation}{section}

\begin{document}

 \pagestyle{empty}

 \vskip-10pt
% {\tt \today}
 \hfill {\tt AEI-2007-148}

\begin{center}

\vspace*{2cm}

\noindent

{\LARGE \textsf{\textbf{Generalized conformal realizations \\[5mm] 
        $\,$of Kac-Moody algebras}} }

\vskip 2truecm

\begin{center}
{\large \textsf{\textbf{Jakob Palmkvist}}} \\
\vskip 1truecm
        {\ffam
        {Albert-Einstein-Institut \\
        Max-Planck-Institut f\"ur Gravitationsphysik\\ 
        Am M\"uhlenberg 1, D-14476 Golm, Germany}\\[3mm]}
        {\tt jakob.palmkvist@aei.mpg.de} \\
\end{center}

\begin{center}
\textit{To the memory of Issai Kantor}
\end{center}

\vskip 1cm

\centerline{\ffam\fser Abstract}

\end{center}

We present a construction which associates an infinite sequence of Kac-Moody algebras, labeled by a positive integer $n$, to one single Jordan algebra. 
For $n=1$, this reduces to the well known Kantor-Koecher-Tits construction. Our generalization utilizes a new relation between different generalized Jordan triple systems, together with their known connections to Jordan and Lie algebras. Applied to the Jordan algebra of hermitian $3\times3$ matrices over the division algebras $\R,\,\C,\,\H,\,\bbO$, the construction gives the exceptional Lie algebras $\mathfrak{f}_{4},\,\frake_{6},\,\frake_{7},\,\frake_{8}$ for $n=2$. Moreover, we obtain their infinite-dimensional extensions for $n \geq 3$. In the case of $2 \times 2$ matrices the resulting Lie algebras are of the form $\so(p+n,\,q+n)$ and the concomitant nonlinear realization generalizes the conformal transformations in a spacetime of signature $(p,\,q)$.

%We present a novel generalization of the Kantor-Koecher-Tits construction, which associates a Lie algebra to any Jordan algebra. This generalizes the conformal transformations in a $(p+q)$-dimensional spacetime to a 
%nonlinear realization of $\so(p+n,\,q+n)$, for arbitrary $n$, with a linearly realized subalgebra $\so(p,\,q)$. 
%For Minkowski spacetimes of $3,\,4,\,6,\,10$ dimensions, the corresponding
%triple systems can be constructed from the Jordan algebras of hermitian 
%$2 \times 2$ matrices over the division algebras $\R,\,\C,\,\H,\,\bbO$, respectively. The construction can also be applied to $3 \times 3$ matrices and then gives rise to the exceptional Lie algebras $\mathfrak{f}_{4},\,\frake_{6},\,\frake_{7},\,\frake_{8}$, as well as to their affine, 
%hyperbolic and further extensions. 
%In particular, this leads to a new realization of the indefinite Kac-Moody algebras $\frake_{10}$ and $\frake_{11}$.

%\newpage

\pagestyle{plain}

\section{Introduction}
Jordan algebras are commutative but non-associtive algebras, which were originally studied in order to understand the foundations of quantum mechanics \cite{Jordan,Jordneuwig}.
Through their connection to Lie algebras, Jordan algebras play an important role in fundamental physics, and 
%Even though the hope of applications to "relativistic and nuclear phenomena" has not been fulfilled, Jordan algebras have turned out to play an important role in fundamental physics through their connection to Lie algebras. In particular, they 
can be used to define generalized spacetimes \cite{Gunaydin:1975mp,Gunaydin7}. 
The origin of this connection lies in the observation that the triple product
\begin{align}
(x,\,y,\,z) \mapsto (xyz) \equiv [[x,\,\tau(y)],\,z]
\end{align}
in the subspace $\g_{-1}$ of a 3-graded Lie algebra $\g_{-1}+\g_0+\g_1$, where $\tau$ is an involution $\g_{-1} \to \g_{1}$,
has the same general properties as the triple product
\begin{align} \label{jtsidentitet}
(x,\,y,\,z) \mapsto (xyz) \equiv  (x y) z+x(y z)-y (x z)
\end{align} 
formed from the multiplication in a Jordan algebra. 
In the Kantor-Koecher-Tits construction \cite{Kantor1,Koecher,Tits1}, any Jordan algebra gives rise to a 3-graded Lie algebra, such that the two triple products coincide.
In the present paper we generalize this construction for a certain kind of Jordan algebras. We will show that any such Jordan algebra gives not only one Lie algebra, but an infinite sequence of Lie algebras, labeled by a positive integer $n$. For $n=1$, we get back the original Kantor-Koecher-Tits construction.

The Kantor-Koecher-Tits construction has already been generalized by Kantor  from Jordan algebras to Jordan triple systems, and further to {generalized} Jordan triple systems \cite{Kantor3.5}.
The generalization of Jordan algebras to Jordan triple systems is needed for the inverse of the Kantor-Koecher-Tits construction -- any 3-graded Lie algebra with an involution $\tau$ gives rise to a Jordan triple system, but not all of them can be obtained from a Jordan algebra by (\ref{jtsidentitet}). Generalized Jordan triple system correspond to graded Lie algebras in general, not necessarily 3-graded. These well known connections between Jordan algebras, (generalized) Jordan triple systems and graded Lie algebras are tools that we will use to derive the results in this paper.

Our construction is based on a new generalization of a single generalized Jordan triple system to an infinite sequence of such triple systems. We study the case when the Lie algebra associated to the first one (the original generalized Jordan triple system) is a finite Kac-Moody algebra, which means that it can be characterized by a Dynkin diagram. We find that continuing the sequence then corresponds to adding more nodes to the Dynkin diagram. Each node will be connected to the previous one by a single line, starting from an arbitrary node in the original Dynkin diagram. For the classical Lie algebras $\mathfrak{b}_r$ ($r \geq 3$) and $\mathfrak{d}_r$ ($r \geq 4$) and for the exceptional Lie algebras $\mathfrak{f}_{4},\,\frake_{6},\,\frake_{7},\,\frake_{8}$, there is a unique %'affine' 
node in the Dynkin diagram such that we get the affine extension if we connect an additional node to it by a single line. In this case our construction
only gives the 
current algebra extension, which means that the central element and the derivation must be added by hand. In all other cases we get the full Kac-Moody algebra, whether it is finite-dimensional or not.

This work is motivated by the 'magic square' constructions \cite{Tits,Vinberg,Kantorartikel,%Sudbery,
Sudbery2},
which associate a Lie algebra $M(\mathbb{K},\,\mathbb{K}')$ to any pair
$(\mathbb{K},\,\mathbb{K}')$ of division algebras $\mathbb{K}=\R,\,\C,\,\mathbb{H},\bbO$. These constructions involve the simple Jordan algebras $H_3(\mathbb{K})$
of hermitian $3 \times 3$
matrices over the division algebras $\mathbb{K}=\R,\,\C,\,\mathbb{H},\bbO$, where the product is the symmetrized matrix product.
Our construction gives a Lie algebra for each simple Jordan algebra and each positive integer value of a parameter $n$ in the following way. The Jordan algebra first leads to a Jordan triple system by (\ref{jtsidentitet}) which in turn generalizes to infinitely many generalized Jordan triple systems. Each of them has an associated Lie algebra. 
We will show that when we apply the construction to the Jordan algebras $H_3(\mathbb{K})$, we obtain the third row in the magic square for $n=1$ (since this is the ordinary Kantor-Koecher-Tits construction) and the fourth row for $n=2$. Moreover, we get the current algebra extension of the algebras in the fourth row for $n=3$ (since the node that we start from is the 'affine' one), and their hyperbolic extensions for $n=4$.   
Thus our construction not only unifies the third and the fourth row, but also extends the magic square with infinitely many new rows. In particular, for $\mathbb{K}=\bbO$, we get a unified construction of $\frake_8,\,\frake_9,\,\frake_{10}$, and further extensions.
When we instead apply our construction to the Jordan algebras $H_2(\mathbb{K})$ 
of hermitian $2 \times 2$
matrices over the division algebras $\mathbb{K}=\R,\,\C,\,\mathbb{H},\bbO$, %where the product is the symmetrized matrix product,
then
the 
associated Lie algebras will always be finite-dimensional, and we consider here not only the complex Lie algebras but also their real forms. In particular, for 
$\mathbb{K}=\bbO$, we get the pseudo-orthogonal algebras $\so(1+n,\,9+n)$, with the conformal algebra in a ten-dimensional Minkowski spacetime as the well known case $n=1$.

Our method is useful for studying nonlinear realizations of Lie algebras
\cite{Gunaydin,Palmkvist:2005gc}.
Any graded Lie algebra can be realized nonlinearly on its 
%Any graded Lie algebra admits a nonlinear realization on its 
subspaces of negative (or positive) degree
\cite{Kantor5}. %When the graded Lie algebra is constructed from a generalized Jordan triple system, then 
This nonlinear realization can be expressed in terms of the corresponding generalized Jordan triple system. When this in turn is obtained from an original one for some $n$ in the way that we will describe, then the nonlinear realization can be expressed in terms of the original generalized Jordan triple system. We will illustrate this for $\so(p+n,\,q+n)$. 

The paper is organized as follows. In Section 2 we show that any generalized Jordan triple system corresponding to a finite Kac-Moody algebra generalizes to an infinite sequence of such triple systems, labeled by a positive integer $n$, and that this corresponds to adding nodes to the original associated Dynkin diagram. In Section 3, we review the relation between Jordan algebras and the magic square of Lie algebras. Then we show that the associated Lie algebras in the $H_3(\mathbb{K})$ case are the exceptional Lie algebras and their extensions.
%, and the pseudo-orthogonal algebras in the $H_2(\mathbb{K})$ case.
The nonlinear realization of $\so(p+n,\,q+n)$, with the linearly realized subalgebra $\so(p,\,q)$ is given in Section 4.
%In Section 4 we first give the nonlinear realization of $\so(p+n,\,q+n)$ with a linearly realized subalgebra $\so(p,\,q)$. Then we show that the special case $n=1$, which corresponds to conformal transformations in a spacetime of signature $(p,\,q)$, is related to the general case in the way it should according to our result. Finally, we go one step further and relate the generalized conformal realization to the Jordan algebras $H_2(\mathbb{K})$ for Minkowski spacetimes in $3,\,4,\,6,\,10$ dimensions.
In the appendix we review in detail how any generalized Jordan triple system gives rise to a graded Lie algebra and how the graded Lie algebra can be nonlinearly realized.

%\newpage

\section{Kac-Moody algebras} %\label{betraktelser}

In this section we will prove our main result (Theorem 2.1) about Kac-Moody algebras and generalized Jordan triple systems. %is proven in Section \ref{betraktelser}.
First we will
briefly recall how a complex Kac-Moody algebra can be constructed from its (generalized) Cartan matrix, or equivalently, from its Dynkin diagram (for details, see \cite{Kac}), and then how it can be given a grading. %Our result 
We will assume that the determinant of the Cartan matrix is non-zero, which in particular means that we leave the affine case for now.
%but we will come back to it in section \ref{sista}.
The Kac-Moody algebra will then be finite-dimensional (or simply \textit{finite}) if and only if the Cartan matrix is positive-definite.
% and infinite-dimensional (or \textit{indefinite})
%if the determinant is negative.  

The Cartan matrix is of type 
$r \times r$, where $r$ is the \textit{rank} of the Lie algebra.
Its entries are integers satisfying
$A^{ii} = 2$ (no summation) and
\begin{align}
i \neq j \Rightarrow A^{ij} \leq 0, \qquad A^{ij} = 0 \Leftrightarrow A^{ji} = 0 
\end{align}
for $i,\,j=1,\,2,\,\ldots,\,r$.
The Dynkin diagram consists of $r$ nodes, and two nodes $i,\,j$ are connected by a line if $A^{ij}=A^{ji}=-1$, but disconnected if $A^{ij}=A^{ji}=0$ (these are the only two cases that we will consider).

In the construction of a Lie algebra from its Cartan matrix, 
one starts with $3r$ generators $e_i,\,f_i,\,h_i$ satisfying the \textit{Chevalley relations}
(no summation)
\begin{align} \label{chevalley-rel}
[e_i,\,f_j] &= \delta_{ij}h_j, & [h_i,\,h_j]&=0,\nn\\
[h_i,\,e_j] &= A_{ij}e_j,& [h_i,\,f_j] &= -A_{ij}f_j.
\end{align}
The elements $h_i$ span the abelian \textit{Cartan subalgebra} $\g_0$.
Further basis elements of $\g$ will then be multiple commutators of either $e_i$ or $f_i$, generated by these elements modulo the \textit{Serre relations}
(no summation)
\begin{align}
({\text{ad }e_i})^{1-A^{ji}}{e_j}&=0, & ({\text{ad }f_i})^{1-A^{ji}}{f_j}&=0.
\end{align}
It follows from (\ref{chevalley-rel}) that these multiple commutators (as well as the elements $e_i$ and $f_i$ themselves) are eigenvectors of $\text{ad }h$ for any $h \in \g_0$, and thus each of them defines an element $\mu$ in the dual space of $\g_0$, such that $\mu(h)$ is the corresponding eigenvalue. These elements $\mu$ are the \textit{roots} of $\g$ and the eigenvectors are called \textit{root vectors}. In particular, $e_i$ are root vectors of the \textit{simple roots} $\alpha_i$, which form a basis of the dual space of $\g_0$. In this basis, an arbitrary root $\mu= \mu^i \alpha_i$ has integer components $\mu^i$, either all non-negative (if $\mu$ is a \textit{positive root}) or all non-positive (if $\mu$ is a \textit{negative root}). 

For finite Kac-Moody algebras, the space of root vectors corresponding to any root is one-dimensional. Furthermore, if $\mu$ is a root, then $-\mu$ is a root as well, but no other multiples of $\mu$. For any positive root $\mu$ of a finite Kac-Moody algebra $\g$, we let $e_\mu$ and $f_{\mu}$ be root vectors corresponding to $\mu$ and $-\mu$, respectively, such that they are multiple commutators of $e_i$ or $f_i$. (This requirement fixes the normalization up to a sign.)
Thus a basis of $\g$ is formed by these root vectors $e_{\mu},\,f_\mu$ for all positive roots $\mu$, and by the Cartan elements $h_i$ for all $i=1,\,2,\,\ldots,\,r$.

\subsection{Graded Lie algebras} \label{gradedlieagebras} 
%\label{gradedlieagebras}

A Lie algebra $\g$ is \textit{graded}, or has a \textit{grading}, if it is the direct sum of subspaces $\g_k \subset \g$ for all integers $k$, such that $[\g_m,\,\g_n] \subset \g_{m+n}$ for all integers $m,\,n$. 
If there is a positive integer $m$ such that $\g_{\pm m} \neq 0$ but $\g_{\pm k} = 0$ for all $k > m$, then the Lie algebra $\g$ is $(2m+1)$-\textit{graded}.
We will occasionally use the notation $\g_{\pm}=\g_{\pm 1}+\g_{\pm 2}+\cdots$.

Any simple root $\alpha_i$ of a Kac-Moody algebra $\g$ \textit{generates} a grading of $\g$, such that for $k\leq0$, the subspace
$\g_k$ ($\g_{-k}$) is spanned by all root vectors $e_{\mu}$ ($f_{\mu}$) with the component $\mu^i = -k$ (the minus sign is a convention) corresponding to $\alpha_i$ in the basis of simple roots, and, if $k=0$, by the Cartan elements $h_j$.

A \textit{graded involution} $\tau$ on the Lie algebra $\g$ is an automorphism such that
$\tau^2(x)=x$ 
for any $x \in \g$ and $\tau(\g_{k}) = \g_{-k}$ for any integer $k$. The simplest example of a graded involution in a graded Kac-Moody algebra is given by $e_{\alpha} \leftrightarrow  \pm f_\alpha$ and $h_i \leftrightarrow -h_i$. (With the minus sign, this is the \textit{Chevalley involution}.) 

On the subspace $\g_{-1}$ of a graded Lie algebra $\g$ with a graded involution $\tau$, we can define a triple product, that is, a trilinear map $(\g_{-1})^3 \to \g_{-1}$, given by
\begin{align}
(x,\,y,\,z) \mapsto (xyz)= [[x,\,\tau(y)],\,z]. \label{standardtrippelprodukten}
\end{align} 
Then, due to the Jacobi identity and the fact that $\tau$ is an involution, this triple product will satisfy the identity
\begin{align}
(uv(xyz))-(xy(uvz))=((uvx)yz)-(x(vuy)z), \label{pre-GJTS-identity}
\end{align}
which means that $\g_{-1}$ is a \textit{generalized Jordan triple system}.
As this name suggests, and as we have already mentioned, this kind of triple systems is related to Jordan algebras. We will explain the relation in more detail in Section \ref{knytaihop}. 

\subsection{Extensions of graded Lie algebras} \label{betraktelser}
We now consider
the situation when a finite Kac-Moody algebra 
$\h$ is extended to another one $\g$ in the following way, for an arbitrary integer $n \geq 2$.

\noindent
%\vspace{-1.3cm}
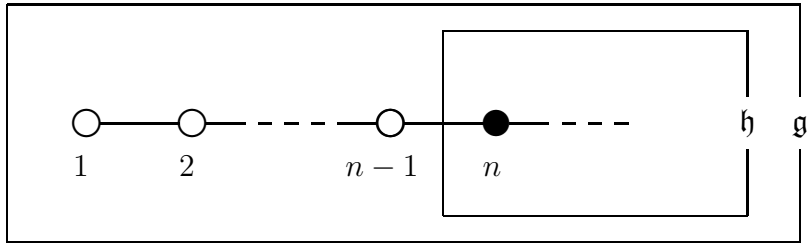
\begin{minipage}{260pt}
\begin{center}
%\scalebox{1}{
\begin{picture}(190,80)(65,-10)
%\put(5,-10){$1$}
%\put(45,-10){$2$}
%\put(85,-10){$3$}
\put(125,-10){$1$}
\put(165,-10){$2$}
%\put(205,-5){$6$}
\put(228,-10){${n-1}$}
\put(280,-10){${n}$}
\put(265,10){\line(0,1){35}}
\put(265,10){\line(0,-1){35}}
\put(380,20){\line(0,1){25}}
\put(377,7){$\h$}
\put(380,0){\line(0,-1){25}}
\put(400,20){\line(0,1){35}}
\put(397,7){$\g$}
\put(400,0){\line(0,-1){35}}
\put(100,10){\line(0,1){45}}
\put(100,10){\line(0,-1){45}}
\put(100,55){\line(1,0){300}}
\put(100,-35){\line(1,0){300}}
\put(265,45){\line(1,0){115}}
\put(265,-25){\line(1,0){115}}
%\put(285,-5){$8$}
%\put(325,-5){$\alpha_{n}$}
%\put(100,45){$n$}
\thicklines
\multiput(130,10)(40,0){2}{\circle{10}}
\multiput(245,10)(40,0){1}{\circle{10}}
\multiput(285,10)(40,0){1}{\circle*{10}}
\multiput(135,10)(40,0){1}{\line(1,0){30}}
\multiput(250,10)(40,0){1}{\line(1,0){30}}
\put(175,10){\line(1,0){15}}
\put(230,10){\line(1,0){10}}
\multiput(195,10)(10,0){4}{\line(1,0){5}}
\put(290,10){\line(1,0){15}}
\multiput(310,10)(10,0){3}{\line(1,0){5}}
\put(245,10){\circle{10}}
%\put(90,50){\circle{10}} \put(90,15){\line(0,1){30}}
\end{picture}
%} 
\end{center}
\vspace*{1cm}
\end{minipage}

\noindent
%The simple root $\alpha_n$ corresponding to the black node $n$, 
%generates a grading of $\g$, where the subspace $\g_k$ is spanned by all root vectors $e_{\mu}$, such that the root $\mu$ has the coefficient $\mu_n=k$ corresponding to $\alpha_n$ in the basis of simple roots. But this is also a simple root of $\h$, so we get a grading of $\h$ in the same way.
The black node,
which $\g$ and $\h$ have in common, generates a grading of $\g$ as well as of 
$\h$. 
We want to investigate how the triple systems $\g_{-1}$ and $\h_{-1}$, corresponding to these two gradings, are related to each other. 
It is clear that $\dim \g_{-1} = n\, \dim \h_{-1}$, which means that $\g_{-1}$ as a vector space is isomorphic to the direct sum $(\h_{-1})^n$ of $n$ vector spaces, each isomorphic to $\h_{-1}$. The question is if we can define a triple product on $(\h_{-1})^n$ such that $\g_{-1}$ and $(\h_{-1})^n$ are isomorphic also as triple systems. To answer this question, we 
write a general element in $(\h_{-1})^{n}$ as $(x_1)^1 + (x_2)^2 + \cdots + (x_{n})^{n}$, where $x_1,\,x_2,\,\ldots$ are elements in $\h_{-1}$.
Furthermore, we define for any graded involution $\tau$ on $\h$ a bilinear form on $\h_{-1}$ 
\textit{associated to} $\tau$
by $(e_{\mu},\,\tau(f_{\nu}))=\delta_{\mu \nu}$ for root vectors
$e_{\mu} \in \h_{-1}$ and $f_{\nu} \in \h_1$.
The answer is then given by the following theorem, which is the main result of this paper.
%(for a proof, see \cite{Palmkvist:2007as}).
\begin{hej} \label{sats1}
The vector space $(\h_{-1})^{n}$, together with the triple product given by
\begin{align}
(x^a y^b z^c)=
\delta^{ab}[[x,\,\tau(y)],\,z]^c  
- \delta^{ab}(x,\,y)z^c
+ \delta^{bc}(x,\,y)z^a
\end{align}
for $a,\,b,\,\ldots=1,\,2,\,\ldots,\,n$ and $x,\,y,\,z \in \h_{-1}$,
is a triple system isomorphic to the triple system $\g_{-1}$ with the triple product 
%\begin{align}
$(uvw)= [[u,\,\tau(v)],\,w]$,
%\end{align}
where the involution $\tau$ is extended from $\mathfrak{h}$ to $\g$ by $\tau(e_i)=-f_i$ for the simple root vectors.
%$(i=1,\,2,\,\ldots,\,\textrm{\rm rank }\h)$.
%Thus $(\h_{-1})^{n}$ is a generalized Jordan triple system, as well as
%$\g_{-1}$.
\end{hej}

\Pf 
%As described in section \ref{gradedlieagebras}, 
A basis of $\h_{-1}$ consists of all root vectors $e_{\mu}$ such that the component of $\mu$ corresponding to $\alpha_{n}$ in the basis of simple roots is equal to one.
A basis of $\g_{-1}$ consists of all such basis elements $e_{\mu}$ of $\h_{-1}$ together with all commutators
$[e^i,\,e_{\mu}]$,
where $e^i$, for $i = 1,\,2,\,\ldots,\,n-1$, is the root vector
\begin{align}
e^i=[\ldots[[e_i,\,e_{i+1}],\,e_{i+2}],\,\ldots,\,e_{n-1}]
\end{align}
of the $\mathfrak{a}_{n-1}$ subalgebra of $\g$. 
We also define the root vector
\begin{align}
f^i=(-1)^{n-1-i}[\ldots[[f_i,\,f_{i+1}],\,f_{i+2}],\,\ldots,\,f_{n-1}]
\end{align}
for the corresponding negative root, and the element
\begin{align}
h^i&=h_i+h_{i+1}+h_{i+2}+\cdots+h_{n-1}
\end{align}
in the Cartan subalgebra of $\g$, such that
\begin{align} \label{basic}
[e^i,\,f^i]=h^i,\qquad
\lbrack h^i,\,e^i]=2e^i,\qquad
\lbrack h^i,\,f^i]=-2f^i,
\end{align}
(no summation). If $i \neq j$, then
\begin{align} \label{feltecken}
\lbrack h^i,\,e^j]=e^j,\qquad
\lbrack h^i,\,f^j]=-f^j,
\end{align}
while $[e^i,\,e^j]$ 
is either zero or a root vector 
of $\g$ that does not belong to $\g_{-1}$.
(We stress the difference between having the indices $i,\,j,\,\ldots$ on
$e,\,f,\,h$ upstairs and downstairs. The root vectors $e_i$ correspond to the simple roots of the $\mathfrak{a}_{n-1}$ subalgebra, while the root vectors 
$e^i$ correspond to roots of 
the $\mathfrak{a}_{n-1}$ subalgebra for which the component corresponding to the simple root $\alpha_{n-1}$ is equal to one, and these roots are not simple, except for $\alpha_{n-1}$ itself.)
Using the relations (\ref{basic})$-$(\ref{feltecken})
we get
\begin{align}
\lbrack [e^i,\,f^j],\,e_{\mu}]&=-\delta^{ij}e_{\mu}, &
\lbrack [e^i,\,f^j],\,f_{\nu}]&=\delta^{ij}f_{\nu},\nn\\
\lbrack [e_{\mu},\,f_{\nu}],\,e^i]&=-\delta_{\mu \nu}e^i,&
\lbrack [e_{\mu},\,f_{\nu}],\,f^j]&=\delta_{\mu \nu}f^j, \label{rel1}
\end{align}
and then
\begin{align}
[[e^i,\,e_{\mu}],\,[f^j,\,f_{\nu}]] = 
-\delta^{ij}[e_{\mu},\,f_{\nu}]-\delta_{\mu\nu}[e^i,\,f^j], \nn
\end{align}
\begin{align}
\lbrack[e^i,\,e_{\mu}],\,f_{\nu}] &= 
\delta_{\mu\nu}e^i, &
\lbrack e_{\mu},\,[f^j,\,f_{\nu}]] &= 
-\delta_{\mu\nu}f^j. \label{rel2}
\end{align}
Finally we have
\begin{align}
\lbrack [e^i,\,f^j],\,e^k] = \delta^{jk}e^i+\delta^{ji}e^k. \label{rel3}
\end{align}
We introduce the bilinear form 
on $\h_{-1}$ \textit{associated to} $\tau$, defined
by $(e_{\mu},\,\tau(f_{\nu}))=\delta_{\mu \nu}$.
Then, from (\ref{rel1})$-$(\ref{rel2}) we get
\begin{align}
\lbrack [e_{\mu},\,\tau(e_{\nu})],\,e^i]&=-(e_{\mu},\, e_{\nu})e^i,&
\lbrack [e_{\mu},\,\tau(e_{\nu})],\,f^j]&=(e_{\mu},\, e_{\nu})f^j, \nn
\end{align}
\begin{align}
[[e^i,\,e_{\mu}],\,[f^j,\,\tau(e_{\nu})]] = 
-\delta^{ij}[e_{\mu},\,\tau(e_{\nu})]-(e_{\mu},\, e_{\nu})[e^i,\,f^j], \nn
\end{align}
\begin{align}
\lbrack[e^i,\,e_{\mu}],\,\tau(e_{\nu})] &= 
(e_{\mu},\, e_{\nu})e^i, &
\lbrack e_{\mu},\,[f^j,\,\tau(e_{\nu})]] &= 
-(e_{\mu},\, e_{\nu})f^j. 
\end{align}
Consider now the direct sum $(\h_{-1})^{n}$ of $n$ vector spaces, each isomorphic to $\h_{-1}$, and write a general element in $(\h_{-1})^{n}$ as $(x_1)^1 + (x_2)^2 + \cdots + (x_{n+1})^{n}$, where $x_1,\,x_2,\,\ldots$ are elements in $\h_{-1}$.
It is easy to see that the map $\psi$, defined by
$e_\mu{}^{1} \mapsto e_{\mu}$ and 
$(e_{\mu})^{i+1} \mapsto [e^{i},\,e_{\mu}]$ for $i = 1,\,2,\,\ldots,\,n-1$
is one-to-one. 
Using the relations above it is straightforward to
show, case by case, that also
%\begin{align}
$\psi((uvw))=(\psi(u)\psi(v)\psi(w))$ %\label{trippelhomo}
%\end{align}
for all $u,\,v,\,w \in (\h_{-1})^{n}$. 
\qed

\section{Jordan algebras and magic squares} \label{Jordan}

In the Kantor-Koecher-Tits construction, any Jordan algebra $J$ 
is associated to a 
%Lie algebra s
%any Jordan triple system gives rise to a 
3-graded Lie algebra $\g_{-1}+\g_0+\g_1$, spanned by the operators %The Lie algebra associated to a Jordan triple system is defined as follows.
%The operators
\begin{align}
%\begin{displaymath}
%$a$ & $b$ & $c$ & $d$
u &\in {\g_{-1}:} &   x&\mapsto u & &{{\text{(constant)}}}\nonumber\\
[u,\,\tau(v)]&\in{\g_{0}:} &   x&\mapsto(uvx) & &{\text{(linear)}}\nonumber\\
\tau(u) &\in {\g_{1}:} &   x&\mapsto-\tfrac{1}{2}(xux) & &{\text{(quadratic)}}
\label{jordanoperatorer}
\end{align}
acting on the Jordan algebra, where
\begin{align}
(xyz)=(xy)z+x(yz)-y(xz).
\end{align}
%where $x$ is an element in the Jordan triple system, which can be identified with $\g_{-1}$.
%acting on a Jordan triple system $J$, where $u,\,v \in J$,
%span the 3-graded Lie algebra \textit{associated} to $J$.
%If the triple system is derived from a Jordan algebra $J$ by
%(\ref{jtsidentitet}),
%\begin{align*}
%(xyz)=(x \circ y) \circ  z+x \circ (y \circ z)-y \circ (x \circ z),
%\end{align*}
The associated Lie algebra $\g_{-1}+\g_0+\g_1$ is 
the \textit{conformal algebra}
$\con \, J$, and $\g_0$ is the \textit{structure algebra} $\str \, J$.
If $J$ has an identity element, then all scalar multiplications form a one-dimensional ideal of $\str\, J$. Factoring out this ideal, we obtain the 
\textit{reduced} structure algebra $\str' J$, which in turn contains the Lie algebra 
$\der \, J$ of all \textit{derivations} of $J$.
%form a subalgebra $\der \, J$ of $\str' J$.
Thus,
%\begin{align*}
$\g_0 = \str \, J \supset \str' J \supset \der \, J$. 
%\end{align*}
%This construction of a 3-graded Lie algebra from a given Jordan algebra is called the {Kantor-Koecher-Tits construction} 
%\cite{Kantor1,Koecher,Tits1}.
%[3,\,5,\,9].
To see why the resulting Lie algebra is called 'conformal' we consider  
the Jordan algebras $H_2(\mathbb{K})$ of hermitian $2 \times 2$
matrices over the division algebras $\mathbb{K}=\R,\,\C,\,\mathbb{H},\bbO$. We have
\begin{align} \nonumber
\der \,  H_2(\mathbb{K})&= \so(d-1)\\ \nonumber
\str'\,  H_2(\mathbb{K})&= \so(1,\,d-1)\\
\con \,  H_2(\mathbb{K})&= \so(2,\,d) %\label{pseudokonstruktion}
\end{align}
for $d=3,\,4,\,6,\,10$, respectively \cite{Sudbery}. It is well known that $\con \,  H_2(\mathbb{K})$ is the algebra that generates conformal transformations in a $d$-dimensional Minkowski spacetime. Furthermore, $\str'\,  H_2(\mathbb{K})$ is the Lorentz algebra and $\der \,  H_2(\mathbb{K})$ its spatial part.

The Kantor-Koecher-Tits construction can be applied also to the Jordan algebras $H_3(\mathbb{K})$ of hermitian $3 \times 3$ matrices over $\mathbb{K}$.
Then we obtain the first three rows in a 'magic square' of Lie algebras \cite{Tits,Vinberg,Kantorartikel,Sudbery2,
Sudbery}. The magic square construction associates a Lie algebra 
$M(\mathbb{K},\,\mathbb{K}')$
to any pair $(\mathbb{K},\,\mathbb{K}')$ of division algebras
$\R,\,\C,\,\H,\,\bbO$,
in a natural way that leads the following
symmetric $4 \times 4$ array.
\\\\\\
\makebox[\textwidth][c]{
\begin{tabular}{|c|c|c|c|c|}

\hline

$\mathbb{K}' \backslash \mathbb{K}$        &  $\R$      &  $\C$                    &  $\H$         &  $\bbO$     \\ \hline
$\R$     &  $\mathfrak{a}_1$  &  $\mathfrak{a}_2$                &  $\mathfrak{c}_3$
    
&  $\f_{4}$     \\ \hline
$\C$     &  $\quad \mathfrak{a}_2 \quad$  &  $\mathfrak{a}_2 \oplus \mathfrak{a}_2$  &  $\quad \mathfrak{a}_5 \quad$     &  $\quad \frake_{6} \quad$ \\ \hline
$\H$     &  $\mathfrak{c}_3$  &  $\mathfrak{a}_5$                &  $\mathfrak{d}_6$    &  $\frake_{7}$ \\

\hline
$\bbO$   &  $\f_{4}$    &  $\frake_{6}$              &  $\frake_{7}$   &  $\frake_{8}$ \\

\hline \multicolumn{5}{c}{\,}\\
\end{tabular}}
\\\\
For simplicity, %we do not write out the expression for the Lie bracket here,
and we only specify the complex Lie algebras here. In this magic square, the real Lie algebras would actually be the compact forms of the complex Lie algebras that we have specified, but we also get other magic squares of real Lie algebras if we replace $\mathbb{K}$ or 
$\mathbb{K}'$ by the corresponding 'split' algebra $\C^s,\,\H^s,\,\bbO^s$
\cite{Mondocavhandling}.
When $\mathbb{K}'$ is split and $\mathbb{K}$ non-split, 
we get 
the derivation, reduced structure and conformal algebras of $H_3(\mathbb{K})$ as the first three rows. %(and in particular (\ref{kkt}) in the last column). 
When $\mathbb{K}$ and $\mathbb{K}'$ are both split, we get the 
split real forms of the complex Lie algebras above.
We focus on the $3 \times 3$ subsquare in the lower right corner, consisting of simply-laced algebras, with the following Dynkin diagrams.
\\
\newcommand{\esex}{
\begin{picture}(70,30)(-5,-5)
\thicklines
\multiput(0,0)(15,0){5}{\circle{5}}
\multiput(2.5,0)(15,0){4}{\line(1,0){10}}
\put(30,2.5){\line(0,1){10}}
\put(30,15){\circle{5}}
\put(30,0){\circle*{5}}
\end{picture}}

\newcommand{\esju}{
\begin{picture}(85,30)(-5,-5)
\thicklines
\multiput(0,0)(15,0){6}{\circle{5}}
\multiput(2.5,0)(15,0){5}{\line(1,0){10}}
\put(45,2.5){\line(0,1){10}}
\put(45,15){\circle{5}}
\put(60,0){\circle*{5}}
\end{picture}}

\newcommand{\eatta}{
\begin{picture}(100,30)(-5,-5)
\thicklines
\multiput(0,0)(15,0){7}{\circle{5}}
\multiput(2.5,0)(15,0){6}{\line(1,0){10}}
\put(60,2.5){\line(0,1){10}}
\put(60,15){\circle{5}}
\put(15,0){\circle*{5}}
\end{picture}}

\newcommand{\esexminus}{
\begin{picture}(70,30)(-5,-5)
\thicklines
\multiput(0,0)(15,0){5}{\circle{5}}
\multiput(2.5,0)(15,0){4}{\line(1,0){10}}
%\put(30,2.5){\line(0,1){10}}
%\put(30,15){\circle{5}}
\put(30,0){\circle*{5}}
\end{picture}}

\newcommand{\esjuminus}{
\begin{picture}(85,30)(-5,-5)
\thicklines
\multiput(0,0)(15,0){5}{\circle{5}}
\multiput(2.5,0)(15,0){4}{\line(1,0){10}}
\put(45,2.5){\line(0,1){10}}
\put(45,15){\circle{5}}
\put(60,0){\circle*{5}}
\end{picture}}

\newcommand{\eattaminus}{
\begin{picture}(100,30)(-5,-5)
\thicklines
\multiput(15,0)(15,0){6}{\circle{5}}
\multiput(17.5,0)(15,0){5}{\line(1,0){10}}
\put(60,2.5){\line(0,1){10}}
\put(60,15){\circle{5}}
\put(15,0){\circle*{5}}
\end{picture}}

\newcommand{\esexmminus}{
\begin{picture}(70,30)(-5,-5)
\thicklines
\multiput(0,0)(15,0){2}{\circle{5}}
\multiput(45,0)(15,0){2}{\circle{5}}
\multiput(2.5,0)(45,0){2}{\line(1,0){10}}
%\put(30,2.5){\line(0,1){10}}
%\put(30,15){\circle{5}}
%\put(30,0){\circle*{5}}
\end{picture}}

\newcommand{\esjumminus}{
\begin{picture}(85,30)(-5,-5)
\thicklines
\multiput(0,0)(15,0){4}{\circle{5}}
\multiput(2.5,0)(15,0){3}{\line(1,0){10}}
\put(45,2.5){\line(0,1){10}}
\put(45,15){\circle{5}}
%\put(60,0){\circle*{5}}
\end{picture}}

\newcommand{\eattamminus}{
\begin{picture}(100,30)(-5,-5)
\thicklines
\multiput(30,0)(15,0){5}{\circle{5}}
\multiput(32.5,0)(15,0){4}{\line(1,0){10}}
\put(60,2.5){\line(0,1){10}}
\put(60,15){\circle{5}}
%\put(15,0){\circle*{5}}
\end{picture}}

\newcommand{\bildC}{
\begin{picture}(30,30)(-5,-5)
\put(5,0){$\C$}
%\put(15,0){\circle*{5}}
\end{picture}}

\newcommand{\bildH}{
\begin{picture}(30,30)(-5,-5)
\put(5,0){$\H$}
%\put(15,0){\circle*{5}}
\end{picture}}

\newcommand{\bildbbO}{
\begin{picture}(30,30)(-5,-5)
\put(5,0){$\bbO$}
%\put(15,0){\circle*{5}}
\end{picture}}

\newcommand{\bildKK}{
\begin{picture}(30,30)(-5,-5)
\put(-2,0){$\mathbb{K}'\backslash \mathbb{K}$}
%\put(15,0){\circle*{5}}
\end{picture}}

\begin{displaymath}
\begin{array}{|c|c|c|c|}
\hline
\bildKK & \bildC & \bildH & \bildbbO \\
\hline 
\bildC &\esexmminus & \esjumminus & \eattamminus \\ \hline
\bildH &\esexminus & \esjuminus & \eattaminus \\ \hline
\bildbbO & \esex & \esju & \eatta \\ \hline
\end{array}
\end{displaymath}
\label{kvadrat-sida}

In the middle row of the $3\times3$ subsquare above, %(the simple root corresponding to) 
the black node generates the 3-grading of the conformal algebra. (Here and below, this meaning of a black node in a Dynkin diagram should not be confused with any different meaning used elsewhere.)
The outermost node next to it in the last row
%black node is represents the simple root that  
generates
%The Dynkin diagrams in the last row are drawn with one black node each.
%The outermost node next to it represents the simple root that generates the 
the unique 5-grading where 
the subspaces $\g_{\pm 2}$ are one-dimensional. %This 5-grading, which is unique, gives rise to the corresponding Freudenthal triple system. 
With this 5-grading, the algebras in the last row are called 'quasiconformal', associated to {Freudenthal triple systems} \cite{Kantor4,Faulkner,Gunaydin}. This is usually the way 
$\frake_8$ is included in the context of Jordan algebras and octonions.
%The approach in this contribution is different: we want to generalize the conformal realization but keep the linear realization of the reduced structure algebra, and therefore we 
Here we are more interested in the grading generated by the black node itself. Then we have the same situation in the last row as in Section \ref{betraktelser}, for $n=2$. With the notation used in Section \ref{betraktelser}
we thus have $\g$ in the last row and $\h$ in the second last row. Theorem 2.1 now implies that the algebras in the last row follow after those in the second last row in the sequence of Lie algebras that is obtained from $H_3(\mathbb{K})$
via the (generalized) Jordan triple systems. (This holds even for $\mathbb{K}=\R$ although we have not included it in the illustration above.) However, this sequence does not end with $n=2$ but can be continued to infinity, with one Kac-Moody algebra for each positive integer value of $n$. Since the black node in
the last row is the 'affine' one, we will get the corresponding current algebra for $n=3$, and the hyperbolic extension for $n=4$. 

If we apply Theorem 2.1 to
the Jordan algebras $H_2(\mathbb{K})$ instead of
$H_3(\mathbb{K})$, then we get 
%the finite Kac-Moody algebras 
$\mathfrak{d}_{r+n}$ where $r=2,\,3,\,5$ for $\mathbb{K}=\C,\,\H,\,\bbO$, respectively. We will show this in detail in the next section.

%(Here we have included a factor 2, which will turn out to be convenient, and denoted the symmetrized matrix product with a circle.)

%(This holds even for $\mathbb{K}=\R$ although we have not included it in the illustration above.)

%However, this sequence does not end with $n=2$ but can be continued to infinity, with one Kac-Moody algebra for each positive integer value of $n$. 

\section{Application to pseudo-orthogonal algebras}

\label{pseudo-kapitel}

In this section we first give the nonlinear realization of $\so(p+n,\,q+n)$ with a linearly realized subalgebra $\so(p,\,q)$. Then we show that the special case $n=1$, which corresponds to conformal transformations in a spacetime of signature $(p,\,q)$, is related to the general case in the way it should, according to Theorem 2.1. Finally, we relate the generalized conformal realization to the Jordan algebras $H_2(\mathbb{K})$ for Minkowski spacetimes in $3,\,4,\,6,\,10$ dimensions.

%We will now apply the general considerations in the preceding section 
%\ref{betraktelser}, and the generalized conformal realization in Section \ref{associated}, 
%to pseudo-orthogonal algebras.
We start with some basic facts.
Let $V$ be a real vector space with an inner product. The real Lie group $SO(V)$ consists of all endomorphisms $F$ of $V$
which preserve the inner product,
\begin{equation}
(F(u),\,F(v)) = (u,\,v)
\end{equation} 
for all $u,\,v \in V$. The corresponding real Lie algebra $\so(V)$ consists of all endomorphisms $f$ of $V$
which are antisymmetric with respect to the inner product,
\beq
(f(u),\,v)+(u,\,f(v))=0
\eeq
for all $u,\,v \in V$. If $V$ is non-degenerate and finite-dimensional with signature $(p,\,q)$, then 
we can identify $SO(V)$ with the real Lie group $SO(p,\,q)$ consisting of all real $(p+q) \times (p+q)$ matrices $X$
such that
\begin{equation}
\quad X^t \eta X = \eta, \quad \det X=1,
\end{equation}
where $\eta$ is the diagonal matrix associated to the inner product.
Correspondingly, we can identify $\so(V)$ with the real Lie algebra $\so(p,\,q)$ consisting of all $(p+q) \times (p+q)$ matrices $x$
such that
\beq
x^t \eta + \eta x =0.
\eeq
In other words, $\so(p,\,q)$ consists of all real matrices of the form
\beq x=
\begin{pmatrix}
a    & c^t \\
c    & b
\end{pmatrix},
\eeq
where 
$a$ and $b$ are orthogonal $p \times p$ and $q \times q$ matrices respectively.
These groups and algebras are said to be 
\textit{pseudo-orthogonal} 
or, if $p=0$, \textit{orthogonal},
written simply $SO(q)$ and $\so(q)$.

We consider now the pseudo-orthogonal algebra $\so(p+n,\,q+n)$, with the inner product given by 
\begin{align}
\eta = 
\text{diag}(\underbrace{ -1,\,\ldots,\,-1 }_{p},\, \underbrace{ +1,\,\ldots,\,+1}_{q},\,\underbrace{ -1,\,\ldots,\,-1}_{n},\,\underbrace{ +1,\,\ldots,\,+1}_{n}),
\end{align}
for some arbitrary positive integers $n,\,p,\,q$.
It is spanned by all matrices ${G^I}_J$, where
the entry in row $L$, 
column $K$ is given by 
\beq
({G^I}_J)^K{}_L = {\delta^I}_L {\delta^K}_J-\eta^{IK}\eta_{JL},
\eeq
and $I,\,J,\,\ldots = 0,\,1,\,\ldots,\,p+q+2n-1$.
It follows that $({G^I}_J)^t={G^J}_I$. 
If $I \neq J$, then the entry of ${G^I}_J$ in row $I$, column $J$ is 1 while the entry in row $J$, column $I$ is $\pm 1$
and all the others are zero. If $I=J$, then ${G^I}_J=0$. 
These matrices satisfy the commutation relations
\begin{align}
[{G^I}_J,\,{G^K}_L] &= 
{\delta^I}_L {G^K}_J-{\delta^K}_J {G^I}_J + \eta^{IK}\eta_{JM} {G^M}_L - \eta_{JL} \eta^{IM} {G^K}_M,
\end{align}
and all those ${G^I}_J$ with $I < J$ (say) form a basis of $\so(p+n,\,q+n)$.

For $\mu,\,\nu,\,\ldots=0,\,1,\,\ldots,\,p+q-1$ and $a,\,b,\,\ldots=1,\,2,\,\ldots,\,n$, with $\mu < \nu$ and $a < b$,
we take the linear combinations
\begin{align} \nonumber
K_{ab} &= \tfrac{1}{2}(-G^{a+m+n}_{b+m+n}+G^{a+m}_{b+m+n}-G^{a+m+n}_{b+m}+G^{a+m}_{b+m}),\\ \nonumber
{K^{\mu}}_a &= -{G^{\mu}}_{a+m+n}-{G^{\mu}}_{a+m},\\ \nonumber
{D^a}_b &= \tfrac{1}{2}(G^{a+m+n}_{b+m+n}+G^{a+m}_{b+m+n}+G^{a+m+n}_{b+m}+G^{a+m}_{b+m}),\\ \nonumber
{P_{\mu}}^a &= -{G_{\mu}}^{a+m+n}-{G_{\mu}}^{a+m},\\
P^{ab} &= \tfrac{1}{2}(-G^{a+m+n}_{b+m+n}-G^{a+m}_{b+m+n}+G^{a+m+n}_{b+m}+G^{a+m}_{b+m}), \label{basen}
\end{align}
as a new basis, where we have set $m=p+q-1$ for convenience.
We note that $K_{ab}$ and $P^{ab}$ vanish when $n=1$, since they are antisymmetric in the indices $a,\,b$.
The basis elements (\ref{basen}) satisfy the commutation relations
\begin{align*}
[{G^{\mu}}{\nu},\,{D^a}_b] = [{G^{\mu}}_{\nu},\,P^{ab}] = [{G^{\mu}}_{\nu},\,K_{ab}] =0,
\end{align*}
\begin{align*}
\lbrack G^{\mu}{}_{\nu},\,G^{\rho}{}_{\sigma}] 
= \delta^{\mu}{}_{\sigma} G^{\rho}{}_{\nu}-\delta^{\rho}{}_{\nu} G^{\mu}_{\sigma} 
+ \eta^{\mu \rho}\eta_{\nu \lambda} G^{\lambda}{}_{\sigma} - \eta_{\nu \sigma} \eta^{\mu \lambda} G^{\rho}{}_{\lambda},
\end{align*}
\begin{align*}
[{D^a}_b,\,{D^c}_d] &= {{\delta}^a}_d {D^c}_b - {{\delta}^c}_b {D^a}_d,
\end{align*}
\begin{align*}
[{P_{\mu}}^a,\,{K^{\nu}}_b] &= 2({\delta^a}_b {G^{\nu}}_{\mu}-{{\delta}^{\nu}}_{\mu}{D^a}_b),
\end{align*}
\begin{align*}
[{G^{\mu}}_{\nu},\,{P_{\rho}}^a] &= {{\delta}^{\mu}}_{\rho} {P_{\nu}}^a - {\eta}_{\nu \rho} {\eta}^{\mu \lambda} {P_{\lambda}}^a, &
[{G^{\mu}}_{\nu},\,{K^{\rho}}_a] &= -{{\delta}^{\rho}}_{\nu} {K^{\mu}}_a + {\eta}^{\mu \rho} {\eta}_{\nu \lambda} {K^{\lambda}}_a,\\ 
[{D^a}_b,\,{P_{\mu}}^c] &= -{{\delta}^c}_b {P_{\mu}}^a, & 
[{D^a}_b,\,{K^{\mu}}_c] &= {{\delta}^a}_c {K^{\mu}}_b, \\
[{P_{\mu}}^a,\,{P_{\nu}}^b] &= -2{\eta}_{\mu \nu}P^{ab}, &
[{K^{\mu}}_a,\,{K^{\nu}}_b] &= -2{\eta}^{\mu \nu}K_{ab}, \\
[P^{ab},\,{P_{\mu}}^c] &= 0, &
[K_{ab},\,{K^{\mu}}_c] &= 0,
\end{align*}
\begin{align*}
[{D^a}_b,\,P^{cd}] &= {{\delta}^d}_b P^{ac} - {{\delta}^c}_b P^{ad}, & [{D^a}_b,\,K_{cd}] &= {{\delta}^a}_c K_{bd} - {{\delta}^a}_d K_{bc},\\
[{K^{\mu}}_a,\,P^{bc}] &= {{\delta}^c}_a {\eta}^{\mu \lambda}{P_{\lambda}}^b - {{\delta}^b}_a {\eta}^{\mu \lambda}{P_{\lambda}}^c, & 
[{P_{\mu}}^a,\,K_{bc}] &= {{\delta}^a}_c {\eta}_{\mu \lambda}{K^{\lambda}}_b - {{\delta}^a}_b {\eta}_{\mu \lambda}{K^{\lambda}}_c,\\
[P^{ab},\,P^{cd}]&=0, & [K_{ab},\,K_{cd}]&=0,
\end{align*}
\begin{align}
[P^{ab},\,K_{cd}] = {\delta^a}_c {D^b}_d - {\delta^b}_c {D^a}_d - {\delta^a}_d {D^b}_c + {\delta^b}_d {D^a}_c. \label{komm-rel}
\end{align}
We see that $\so(p+n,\,q+n)$ has the following 5-grading, which reduces to a 3-grading when $n=1$.
\\\\
\makebox[\textwidth][c]{

\begin{tabular}{l|c|c|c|c|c}

subspace$\quad$ & $\ \g_{-2}\ $ & $\ \g_{-1}\ $ & $\ \g_{0}\ $ & $\ \g_{1}\ $ & $\ \g_{2}\ $ \\ &&&& \\
\hline &&&& \\
basis & $\ P^{ab}\ $ & $\ {P_{\mu}}^a\ $ & $\ {G^{\mu}}_{\nu},\,{D^a}_b\ $ & $\ {K^{\mu}}_a\ $ & $\ K_{ab}\ $
\end{tabular}}
\\\\\\ 
Furthermore, we see that ${D^a}_b$ satisfy the commutation relations for $\mathfrak{gl}(n,\,\mathbb{R})$. Since they also commute with ${G^{\mu}}_{\nu}$,
we have
%\begin{align}
${\g}_0=\so(p,\,q) \oplus \mathfrak{gl}(n,\,\R)$
%\end{align}
as a direct sum of subalgebras.
Finally, a graded involution $\tau$ is given by
\begin{align} \label{gardin}
\tau({P_{\mu}}^a) &= {\eta}_{\mu \nu}{K^{\nu}}_a = -{G_{\mu}}^{a+m+n}+{G_{\mu}}^{a+m},\nn\\\nn\\
%\end{align}
%and it follows that
%\begin{align}
\tau({K^{\mu}}_a) &= {\eta}^{\mu \nu}{P_{\nu}}^a = -{G^{\mu}}_{a+m+n}+{G^{\mu}}_{a+m},\nn
\end{align}
\begin{align}
\tau({D^a}_b) &= -{D^b}_a,
& \tau(K_{ab}) &= P^{ab},\nn\\
\tau({G^{\mu}}_{\nu}) &= {G^{\mu}}_{\nu} & \tau(P^{ab}) &= K_{ab}.
\end{align}
Thus 
%all conditions are satisfied for 
$\g_{-1}$ 
%to be a Kantor triple system $K$ 
is a generalized Jordan triple system
with the triple product
\begin{align} 
({P_{\mu}}^a {P_{\nu}}^b {P_{\rho}}^c)&= \nonumber
[[{P_{\mu}}^a,\,\tau({P_{\nu}}^b)],\,{P_{\rho}}^c] = 
[[{P_{\mu}}^a,\,{\eta}_{\nu \lambda}{K^{\lambda}}_b],\,{P_{\rho}}^c]\\ \nonumber
&= -2{\delta}^{ab}{\eta}_{\mu \lambda}({{\delta}^{\lambda}}_{\rho}{P_{\nu}}^c-
{\eta}_{\nu \rho}{\eta}^{\lambda \kappa}{P_{\kappa}}^c)
+2{\delta}^{bc}{\eta}_{\mu \nu}{P_{\rho}}^a\\ \label{trippelprodukten}
&= 2{\delta}^{ab}({\eta}_{\nu \rho}{P_{\mu}}^c-{\eta}_{\mu \rho}{P_{\nu}}^c)+2 {\delta}^{bc}{\eta}_{\mu \nu}{P_{\rho}}^a. 
\end{align}
If we now insert (\ref{trippelprodukten}) in (\ref{kantoroperatorer}) (but rescale the elements in $\g_{-2}$ according to \cite{Palmkvist:2005gc}), and
use the isomorphism (\ref{vektorfaltisomomorfi}), so that we 
identify any operator $f$
with the vector field
$-{f^{\mu}}_a {\partial_{\mu}}^a - f_{ab} \partial^{ab}$,
then we get the realization
\begin{align} \nonumber
P^{ab} &= -2{\partial}^{ab},\\ \nonumber
{P_{\mu}}^a &= {\partial_{\mu}}^a-2{x_{\mu b}}{\partial}^{ab},\\ \nonumber
{G^{\mu}}_{\nu} &= {x_{\nu a}}{\partial^{\mu a}}-{x^{\mu}}_a{\partial_{\nu}}^a,\\ \nonumber
{D^a}_b &= {x^{\mu}}_b{\partial_{\mu}}^a+2x_{bc}\partial^{ac},\\ \nonumber
{K^{\mu}}_a &= -2{x^{\nu}}_a {x^{\mu}}_b {\partial_{\nu}}^b + 
{x^{\nu}}_a {x_{\nu b}} {\partial^{\mu b}}
-x_{ab} {\partial^{\mu b}}\\ \nonumber
            & \quad -2{x^{\nu}}_a {x^{\mu}}_b {x_{\nu c}} \partial^{bc} + 2x_{ab}{x^{\mu}}_c \partial^{bc},\\ \nonumber
K_{ab} &= {x^{\mu}}_a {x^{\nu}}_b {x_{\mu c}} {\partial_{\nu}}^c
-{x^{\mu}}_b {x^{\nu}}_a {x_{\mu c}} {\partial_{\nu}}^c - x_{ac} {x^{\mu}}_b {\partial_{\mu}}^c+ x_{bc} {x^{\mu}}_a {\partial_{\mu}}^c\\
            & \quad +
2{x^{\mu}}_a {x^{\nu}}_b {x_{\mu c}} {x_{\nu d}} \partial^{cd} - 2 x_{ac}x_{bd}\partial^{cd}.
\end{align}
Straightforward calculations show that these generators indeed satisfy the commutation relations (\ref{komm-rel}). When $n=1$, 
the $\mathfrak{gl}(n,\,\R)$ indices $a,\,b,\,\,\ldots$ take only one value, so we can suppress them, and everything antisymmetric in these indices vanishes.
We are then left with the conformal realization:
%the usual conformal realization (\ref{konfrel}).
\begin{align}  \nonumber
{P_{\mu}} &= {\partial_{\mu}} &&{\text{(translations)}}  \\
{G^{\mu}}_{\nu} &= %{\eta}^{\mu \rho}{\eta}_{\nu \sigma} \nonumber
{x_{\nu}}{\partial^{\mu}}-{x^{\mu}}{\partial_{\nu}} &&{\text{(Lorentz transformations)}}\nonumber \\ \nonumber
{D} &= {x^{\mu}}{\partial_{\mu}} &&{\text{(dilatations)}}  \\
{K^{\mu}} &= -2{x^{\nu}} {x^{\mu}} {\partial_{\nu}} + 
%{\eta}^{\mu \nu}{\eta}_{\rho \sigma}
{x^{\nu}} {x_{\nu}} {\partial^{\mu}} &&{\text{(special conformal
%$\ldots$
transformations)}}%\\ & &&{\quad \text{transformations)
%$\ldots$
%}} 
\label{konfrel}
\end{align}

We will now show that the 5-grading of $\so(p+n,\,q+n)$ in this section is generated (as described in Section \ref{gradedlieagebras}) by the simple root corresponding to node $n$ in the Dynkin diagram below of $\mathfrak{d}_{r}$ for $p+q=2r$. We will then show that the cases $n=1$ and $n>1$ are related in the way that we described in Section \ref{betraktelser}.\\
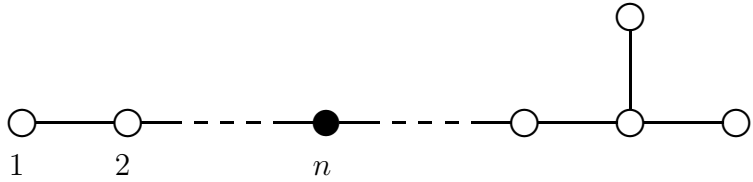
\begin{minipage}{260pt}
\begin{center}
\scalebox{1}{
\begin{picture}(300,80)(65,-10)
%\put(5,-10){$1$}
%\put(45,-10){$2$}
%\put(85,-10){$3$}
\put(125,-10){$1$}
\put(165,-10){$2$}
%\put(205,-5){$6$}
\put(240,-10){${n}$}
\thicklines
\multiput(130,10)(40,0){2}{\circle{10}}
\multiput(320,10)(75,0){1}{\circle{10}}
\multiput(135,10)(40,0){1}{\line(1,0){30}}
\multiput(250,10)(40,0){0}{\line(1,0){30}}
\put(175,10){\line(1,0){15}}
\put(230,10){\line(1,0){10}}
\multiput(195,10)(10,0){4}{\line(1,0){5}}
\multiput(250,10)(50,0){2}{\line(1,0){15}}
\multiput(270,10)(10,0){3}{\line(1,0){5}}
\multiput(360,10)(40,0){1}{\circle{10}}
\put(360,50){\circle{10}}
\put(400,10){\circle{10}}
\put(245,10){\circle*{10}}
\put(365,10){\line(1,0){30}}
\put(325,10){\line(1,0){30}}
\put(360,15){\line(0,1){30}}
%\put(90,50){\circle{10}} \put(90,15){\line(0,1){30}}
\end{picture}}\end{center}
\vspace*{0.1cm}  %{1.8cm}
\end{minipage}
\label{dynkin-sida}\\
\noindent
For this we must
relate the $P_{\mu}$ basis of $\g_{-1}$ used in this section to the basis consisting of root vectors. The relation will of course be different for different Lie algebras $\so(p+n,\,q+n)$. We consider first the case $n=1$. Below we give explicitly the relations (with a suitable choice of Cartan-Weyl generators) for two examples, $(p,\,q)=(5,\,5)$ 
in the left table and 
$(p,\,q)=(1,\,9)$ 
in the right table.
\begin{center}
\begin{tabular}{|c|c|c|}
\hline
$\mu$ & $e_{\mu}$ & $f_{\mu}$ \\
\hline
\begin{picture}(25,20)(0,-3)
\put(0,0){${\scriptstyle 1}$}
\put(5,0){${\scriptstyle 0}$}
\put(10,0){${\scriptstyle 0}$}
\put(15,0){${\scriptstyle 0}$}
\put(20,0){${\scriptstyle 0}$}
\put(15,7){${\scriptstyle 0}$}
\end{picture} &
\begin{picture}(55,20)(0,-5)
\put(0,0){$\tfrac{1}{2}(P_5+P_0)$} 
\end{picture}& 
\begin{picture}(60,20)(0,-5)
\put(0,0){$\tfrac{1}{2}(K^5+K^0)$} 
\end{picture}\\
\begin{picture}(25,20)(0,-3)
\put(0,0){${\scriptstyle 1}$}
\put(5,0){${\scriptstyle 1}$}
\put(10,0){${\scriptstyle 0}$}
\put(15,0){${\scriptstyle 0}$}
\put(20,0){${\scriptstyle 0}$}
\put(15,7){${\scriptstyle 0}$}
\end{picture} &
\begin{picture}(55,20)(0,-5)
\put(0,0){$\tfrac{1}{2}(P_6+P_1)$} 
\end{picture}& 
\begin{picture}(60,20)(0,-5)
\put(0,0){$\tfrac{1}{2}(K^6+K^1)$} 
\end{picture}\\

\begin{picture}(25,20)(0,-3)
\put(0,0){${\scriptstyle 1}$}
\put(5,0){${\scriptstyle 1}$}
\put(10,0){${\scriptstyle 1}$}
\put(15,0){${\scriptstyle 0}$}
\put(20,0){${\scriptstyle 0}$}
\put(15,7){${\scriptstyle 0}$}
\end{picture} &
\begin{picture}(55,20)(0,-5)
\put(0,0){$\tfrac{1}{2}(P_7+P_2)$} 
\end{picture}& 
\begin{picture}(60,20)(0,-5)
\put(0,0){$\tfrac{1}{2}(K^7+K^2)$} 
\end{picture}\\
\begin{picture}(25,20)(0,-3)
\put(0,0){${\scriptstyle 1}$}
\put(5,0){${\scriptstyle 1}$}
\put(10,0){${\scriptstyle 1}$}
\put(15,0){${\scriptstyle 1}$}
\put(20,0){${\scriptstyle 0}$}
\put(15,7){${\scriptstyle 0}$}
\end{picture} &
\begin{picture}(55,20)(0,-5)
\put(0,0){$\tfrac{1}{2}(P_8+P_3)$} 
\end{picture}& 
\begin{picture}(60,20)(0,-5)
\put(0,0){$\tfrac{1}{2}(K^8+K^3)$} 
\end{picture}\\
\begin{picture}(25,20)(0,-3)
\put(0,0){${\scriptstyle 1}$}
\put(5,0){${\scriptstyle 1}$}
\put(10,0){${\scriptstyle 1}$}
\put(15,0){${\scriptstyle 1}$}
\put(20,0){${\scriptstyle 1}$}
\put(15,7){${\scriptstyle 0}$}
\end{picture} &
\begin{picture}(55,20)(0,-5)
\put(0,0){$\tfrac{1}{2}(P_9+P_4)$} 
\end{picture}& 
\begin{picture}(60,20)(0,-5)
\put(0,0){$\tfrac{1}{2}(K^9+K^4)$} 
\end{picture}\\
\begin{picture}(25,20)(0,-3)
\put(0,0){${\scriptstyle 1}$}
\put(5,0){${\scriptstyle 1}$}
\put(10,0){${\scriptstyle 1}$}
\put(15,0){${\scriptstyle 1}$}
\put(20,0){${\scriptstyle 0}$}
\put(15,7){${\scriptstyle 1}$}
\end{picture} &
\begin{picture}(55,20)(0,-5)
\put(0,0){$\tfrac{1}{2}(P_9-P_4)$} 
\end{picture}& 
\begin{picture}(60,20)(0,-5)
\put(0,0){$\tfrac{1}{2}(K^9-K^4)$} 
\end{picture}\\
\begin{picture}(25,20)(0,-3)
\put(0,0){${\scriptstyle 1}$}
\put(5,0){${\scriptstyle 1}$}
\put(10,0){${\scriptstyle 1}$}
\put(15,0){${\scriptstyle 1}$}
\put(20,0){${\scriptstyle 1}$}
\put(15,7){${\scriptstyle 1}$}
\end{picture} &
\begin{picture}(55,20)(0,-5)
\put(0,0){$\tfrac{1}{2}(P_3-P_8)$} 
\end{picture}& 
\begin{picture}(60,20)(0,-5)
\put(0,0){$\tfrac{1}{2}(K^3-K^8)$} 
\end{picture}\\

\begin{picture}(25,20)(0,-3)
\put(0,0){${\scriptstyle 1}$}
\put(5,0){${\scriptstyle 1}$}
\put(10,0){${\scriptstyle 1}$}
\put(15,0){${\scriptstyle 2}$}
\put(20,0){${\scriptstyle 1}$}
\put(15,7){${\scriptstyle 1}$}
\end{picture} &
\begin{picture}(55,20)(0,-5)
\put(0,0){$\tfrac{1}{2}(P_7-P_2)$} 
\end{picture}& 
\begin{picture}(60,20)(0,-5)
\put(0,0){$\tfrac{1}{2}(K^7-K^2)$} 
\end{picture}\\
\begin{picture}(25,20)(0,-3)
\put(0,0){${\scriptstyle 1}$}
\put(5,0){${\scriptstyle 1}$}
\put(10,0){${\scriptstyle 2}$}
\put(15,0){${\scriptstyle 2}$}
\put(20,0){${\scriptstyle 1}$}
\put(15,7){${\scriptstyle 1}$}
\end{picture} &
\begin{picture}(55,20)(0,-5)
\put(0,0){$\tfrac{1}{2}(P_1-P_6)$} 
\end{picture}& 
\begin{picture}(60,20)(0,-5)
\put(0,0){$\tfrac{1}{2}(K^1-K^6)$} 
\end{picture}\\

\begin{picture}(25,20)(0,-3)
\put(0,0){${\scriptstyle 1}$}
\put(5,0){${\scriptstyle 2}$}
\put(10,0){${\scriptstyle 2}$}
\put(15,0){${\scriptstyle 2}$}
\put(20,0){${\scriptstyle 1}$}
\put(15,7){${\scriptstyle 1}$}
\end{picture} &
\begin{picture}(55,20)(0,-5)
\put(0,0){$\tfrac{1}{2}(P_5-P_0)$} 
\end{picture}& 
\begin{picture}(60,20)(0,-5)
\put(0,0){$\tfrac{1}{2}(K^5-K^0)$} 
\end{picture}\\

\hline
\end{tabular} \quad
\begin{tabular}{|c|c|c|}
\hline
$\mu$ & $e_{\mu}$ & $f_{\mu}$ \\
\hline
\begin{picture}(25,20)(0,-3)
\put(0,0){${\scriptstyle 1}$}
\put(5,0){${\scriptstyle 0}$}
\put(10,0){${\scriptstyle 0}$}
\put(15,0){${\scriptstyle 0}$}
\put(20,0){${\scriptstyle 0}$}
\put(15,7){${\scriptstyle 0}$}
\end{picture} &
\begin{picture}(60,20)(0,-5)
\put(0,0){$\tfrac{1}{2}(P_5+P_0)$} 
\end{picture}& 
\begin{picture}(65,20)(0,-5)
\put(0,0){$\tfrac{1}{2}(K^5+K^0)$} 
\end{picture}\\
\begin{picture}(25,20)(0,-3)
\put(0,0){${\scriptstyle 1}$}
\put(5,0){${\scriptstyle 1}$}
\put(10,0){${\scriptstyle 0}$}
\put(15,0){${\scriptstyle 0}$}
\put(20,0){${\scriptstyle 0}$}
\put(15,7){${\scriptstyle 0}$}
\end{picture} &
\begin{picture}(60,20)(0,-5)
\put(0,0){$\tfrac{1}{2}(P_6-\upi P_1)$} 
\end{picture}& 
\begin{picture}(65,20)(0,-5)
\put(0,0){$\tfrac{1}{2}(K^6+\upi K^1)$} 
\end{picture}\\

\begin{picture}(25,20)(0,-3)
\put(0,0){${\scriptstyle 1}$}
\put(5,0){${\scriptstyle 1}$}
\put(10,0){${\scriptstyle 1}$}
\put(15,0){${\scriptstyle 0}$}
\put(20,0){${\scriptstyle 0}$}
\put(15,7){${\scriptstyle 0}$}
\end{picture} &
\begin{picture}(60,20)(0,-5)
\put(0,0){$\tfrac{1}{2}(P_7-\upi P_2)$} 
\end{picture}& 
\begin{picture}(65,20)(0,-5)
\put(0,0){$\tfrac{1}{2}(K^7+\upi K^2)$} 
\end{picture}\\
\begin{picture}(25,20)(0,-3)
\put(0,0){${\scriptstyle 1}$}
\put(5,0){${\scriptstyle 1}$}
\put(10,0){${\scriptstyle 1}$}
\put(15,0){${\scriptstyle 1}$}
\put(20,0){${\scriptstyle 0}$}
\put(15,7){${\scriptstyle 0}$}
\end{picture} &
\begin{picture}(60,20)(0,-5)
\put(0,0){$\tfrac{1}{2}(P_8-\upi P_3)$} 
\end{picture}& 
\begin{picture}(65,20)(0,-5)
\put(0,0){$\tfrac{1}{2}(K^8-\upi K^3)$} 
\end{picture}\\
\begin{picture}(25,20)(0,-3)
\put(0,0){${\scriptstyle 1}$}
\put(5,0){${\scriptstyle 1}$}
\put(10,0){${\scriptstyle 1}$}
\put(15,0){${\scriptstyle 1}$}
\put(20,0){${\scriptstyle 1}$}
\put(15,7){${\scriptstyle 0}$}
\end{picture} &
\begin{picture}(60,20)(0,-5)
\put(0,0){$\tfrac{1}{2}(P_9-\upi P_4)$} 
\end{picture}& 
\begin{picture}(65,20)(0,-5)
\put(0,0){$\tfrac{1}{2}(K^9+\upi K^4)$} 
\end{picture}\\
\begin{picture}(25,20)(0,-3)
\put(0,0){${\scriptstyle 1}$}
\put(5,0){${\scriptstyle 1}$}
\put(10,0){${\scriptstyle 1}$}
\put(15,0){${\scriptstyle 1}$}
\put(20,0){${\scriptstyle 0}$}
\put(15,7){${\scriptstyle 1}$}
\end{picture} &
\begin{picture}(60,20)(0,-5)
\put(0,0){$\tfrac{1}{2}(P_9+\upi P_4)$} 
\end{picture}& 
\begin{picture}(65,20)(0,-5)
\put(0,0){$\tfrac{1}{2}(K^9-\upi K^4)$} 
\end{picture}\\
\begin{picture}(25,20)(0,-3)
\put(0,0){${\scriptstyle 1}$}
\put(5,0){${\scriptstyle 1}$}
\put(10,0){${\scriptstyle 1}$}
\put(15,0){${\scriptstyle 1}$}
\put(20,0){${\scriptstyle 1}$}
\put(15,7){${\scriptstyle 1}$}
\end{picture} &
\begin{picture}(60,20)(0,-5)
\put(0,0){$\tfrac{1}{2}(\upi P_3+P_8)$} 
\end{picture}& 
\begin{picture}(65,20)(0,-5)
\put(0,0){$\tfrac{1}{2}(\upi K^3-K^8)$} 
\end{picture}\\

\begin{picture}(25,20)(0,-3)
\put(0,0){${\scriptstyle 1}$}
\put(5,0){${\scriptstyle 1}$}
\put(10,0){${\scriptstyle 1}$}
\put(15,0){${\scriptstyle 2}$}
\put(20,0){${\scriptstyle 1}$}
\put(15,7){${\scriptstyle 1}$}
\end{picture} &
\begin{picture}(60,20)(0,-5)
\put(0,0){$\tfrac{1}{2}(P_7+\upi P_2)$} 
\end{picture}& 
\begin{picture}(65,20)(0,-5)
\put(0,0){$\tfrac{1}{2}(K^7- \upi K^2)$} 
\end{picture}\\
\begin{picture}(25,20)(0,-3)
\put(0,0){${\scriptstyle 1}$}
\put(5,0){${\scriptstyle 1}$}
\put(10,0){${\scriptstyle 2}$}
\put(15,0){${\scriptstyle 2}$}
\put(20,0){${\scriptstyle 1}$}
\put(15,7){${\scriptstyle 1}$}
\end{picture} &
\begin{picture}(60,20)(0,-5)
\put(0,0){$\tfrac{1}{2}(\upi P_1+P_6)$} 
\end{picture}& 
\begin{picture}(65,20)(0,-5)
\put(0,0){$\tfrac{1}{2}(\upi K^1-K^6)$} 
\end{picture}\\
\begin{picture}(25,20)(0,-3)
\put(0,0){${\scriptstyle 1}$}
\put(5,0){${\scriptstyle 2}$}
\put(10,0){${\scriptstyle 2}$}
\put(15,0){${\scriptstyle 2}$}
\put(20,0){${\scriptstyle 1}$}
\put(15,7){${\scriptstyle 1}$}
\end{picture} &
\begin{picture}(60,20)(0,-5)
\put(0,0){$\tfrac{1}{2}(P_5-P_0)$} 
\end{picture}& 
\begin{picture}(65,20)(0,-5)
\put(0,0){$\tfrac{1}{2}(K^5-K^0)$}
\end{picture}\\
\hline
\end{tabular}
\end{center}
\label{tabell-sida}

\noindent
We have indicated the roots by their coefficients in the basis of simple roots, corresponding to the nodes in the Dynkin diagram above.
(For example, $e_{\mu}=[e_1,\,e_2]$ in the second row, and 
$e_{\mu}=[[e_1,\,e_2],\,e_3]$ in the third.)
It is evident from these tables how to generalize them to arbitrary values of $p,\,q$ (with $p+q$ even).
It follows that the bilinear form associated to the 
involution 
\begin{align} \label{nils}
\tau(P_{\mu}) = \eta_{\mu \nu} K^{\nu}, 
\end{align}
is given in the $P_{\mu}$ basis by $(P_{\mu},\,P_{\nu})=2\eta_{\mu \nu}$.

When we extend $\so(p+1,\,q+1)$ to $\g=\so(p+n,\,q+n)$ for $n > 1$, we put a superscript 1 on $P_{\mu}$ in the expressions for the root vectors $e_{\mu}$ of the subalgebra $\h=\so(p+1,\,q+1)$ in $\h_{-1}$ 
(and a subscript 1 on $K^{\mu}$). 
%In the notation that we used in Section \ref{betraktelser}, we have furthermore $e^a = D^a{}_1$ and $f^a = D^1{}_a$. 
Then the graded involution (\ref{gardin}) on $\g$ is indeed the extension of the original 
involution (\ref{nils}) on $\h$ that we described in Theorem \ref{sats1} and
we get
\begin{align}
({P_{\mu}}^a {P_{\nu}}^b {P_{\rho}}^c) \nonumber
&= 2{\delta}^{ab}{\eta}_{\nu \rho}{P_{\mu}}^c-2{\delta}^{ab}{\eta}_{\mu \rho}{P_{\nu}}^c+2 {\delta}^{bc}{\eta}_{\mu \nu}{P_{\rho}}^a\nn\\
&= 2{\delta}^{ab}{\eta}_{\nu \rho}{P_{\mu}}^c-2{\delta}^{ab}{\eta}_{\mu \rho}{P_{\nu}}^c+2 {\delta}^{ab}{\eta}_{\mu \nu}{P_{\rho}}^c
-2 {\delta}^{ab}{\eta}_{\mu \nu}{P_{\rho}}^c
+2 {\delta}^{bc}{\eta}_{\mu \nu}{P_{\rho}}^a\nn\\
&= {\delta}^{ab}({P_{\mu}} {P_{\nu}} {P_{\rho}})^c
-2 {\delta}^{ab}{\eta}_{\mu \nu}{P_{\rho}}^c
+2 {\delta}^{bc}{\eta}_{\mu \nu}{P_{\rho}}^a \nn\\
&= {\delta}^{ab}[[P_{\mu},\,\tau(P_{\nu})], {P_{\rho}}]^c
- {\delta}^{ab}(P_{\mu},\, P_{\nu}){P_{\rho}}^c
+ {\delta}^{bc}(P_{\mu},\, P_{\nu}){P_{\rho}}^a
\end{align} 
as we should, according to Theorem \ref{sats1}. 

\subsection{Connection to Jordan algebras} \label{knytaihop}

\noindent
When $n=1$, the triple product (\ref{trippelprodukten}) becomes
\begin{align} \label{nlikamedett}
(P_{\mu} P_{\nu} P_{\rho}) =
2{\eta}_{\nu \rho}{P_{\mu}}-2{\eta}_{\mu \rho}{P_{\nu}}+2 {\eta}_{\mu \nu}{P_{\rho}}. 
\end{align}
If we introduce an inner product in the vector space $\g_{-1}$ by 
$P_{\mu}\cdot P_{\nu}=\eta_{\mu \nu}$, then this can be written 
\begin{align}  
(xyz) = 2(z\cdot y)x - 2(z\cdot x)y + 2(x\cdot y)z. \label{komma}
\end{align}
Let $U$ be the subspace of $\g_{-1}$ spanned by $P_i$ for 
$i=1,\,2,\,\ldots,\,p+q-1$. Then we can consider $\g_{-1}$ as the Jordan algebra 
$J(U)$, %(defined in section 2 for Euclidean spaces $U$), 
with the 
product
\begin{align}
P_i \circ P_j = (P_i\cdot P_j)P_0
\end{align} 
for $i,\,j=1,\,2,\,\ldots,\,p+q-1$, and $P_0$ as identity element.
If we introduce a linear map 
\begin{align}
J(U) \to J(U), \quad z \mapsto \tilde{z},
\end{align}
which changes sign on $P_{0}$, but otherwise leaves the basis elements $P_{\mu}$
unchanged, then we can write the inner product as
\begin{align}
2(u \cdot v)=u \circ \tilde{v} + v \circ \tilde{u}. \label{produktrelation}
\end{align}
Inserting (\ref{produktrelation}) in (\ref{komma}), we get
\begin{align} 
(xyz)&=   \nonumber  
 (\tilde{z} \circ {y}) \circ x - (\tilde{z} \circ {x}) \circ y + (\tilde{x} \circ {y})  \circ z \nn\\ 
                 &\quad  + (z \circ {\tilde{y}}) \circ x - (z \circ {\tilde{x}}) \circ y + (x \circ {\tilde{y}}) \circ z\nn\\
                 &=[y,\,\tilde{z},\,x]+[y,\,\tilde{x},\,z]
                 +(z \circ {\tilde{y}}) \circ x + (x \circ {\tilde{y}}) \circ z.
\label{langtuttryck}
\end{align}
It is easy to see that the \textit{associators} in the last line remain unchanged if we move the tilde from one element to another. Thus we get
\begin{align}
(xyz)&= [y,\,\tilde{z},\,x]+[y,\,\tilde{x},\,z]
                 +(z \circ {\tilde{y}}) \circ x + (x \circ {\tilde{y}}) \circ z
 \nn\\    &= [\tilde{y},\,z,\,x]+[\tilde{y},\,x,\,z]
                 +(z \circ {\tilde{y}}) \circ x + (x \circ {\tilde{y}}) \circ z
\nn\\    &= 
2(z \circ \tilde{y}) \circ x - 2(z \circ x) \circ \tilde{y} + 
2(x \circ \tilde{y}) \circ z .\label{enklaformen}
\end{align}
If we instead use the involution given by 
\begin{align} \label{tilde}
\tau(\tilde{P}{}_{\mu}) = \eta_{\mu \nu}{K}^{\nu},
\end{align}
then we can remove the tilde,
\begin{align}
(xyz) = 2(z \circ {y}) \circ x 
- 2(z \circ x) \circ {y} + 2(x \circ {y}) \circ z .
\end{align}
Any Jordan algebra $J$ is also a Jordan triple system with this triple product. The associated Lie algebra, defined by the construction in 
the appendix or in (\ref{jordanoperatorer}),
%Section \ref{Associated}, 
%or in this case equivalently by the realization (\ref{konfrel}), 
is its conformal algebra $\con{J}$.

Consider now the case $p=1$ (and still $n=1$).
Then $U$ is a Euclidean space, and $J(U)$ is a \textit{formally real} 
Jordan algebra.
For $q=2,\,3,\,5,\,9$, there is an isomorphism from $J(U)$ to 
$H_2(\mathbb{K})$,
where $\mathbb{K}=\R,\,\C,\,\H,\,\bbO$, respectively, given by
\begin{align} 
P_0 &\mapsto
\begin{pmatrix} 
1 & 0 \\
0 & 1 
\end{pmatrix},
& P_1 &\mapsto 
\begin{pmatrix} 
0 & 1 \\
1 & 0 
\end{pmatrix},
& P_{i+1} &\mapsto 
\begin{pmatrix} 
0 & -e_{i} \\
e_{i} & 0 
\end{pmatrix},
& P_{p+q-1} &\mapsto 
\begin{pmatrix} 
1 & 0 \\
0 & -1 
\end{pmatrix} \label{isomorfismen}
\end{align}
for $i=1,\,2,\,\ldots,\,p+q-3$. (Here, $e_i$ are the 'imaginary units' that anticommute and square to $-1$.) 
The involution (\ref{tilde}) becomes
%\begin{align}
$\tau({P}{}_{\mu}) = {K}^{\mu}$
%\end{align}
and we see from the tables %on page \pageref{tabell-sida} 
that the associated  bilinear form on $\g_{-1}$
has the simple
form $(x,\,y)=\tr{(x \circ y)}$.
If we instead consider the split form $(p=q)$
for $p=3,\,5,\,9$, then (\ref{isomorfismen}) is still an isomorphism,
if we replace 
$\mathbb{K}=\C,\,\H,\,\bbO$ by the 'split' algebra $\mathbb{K}^s$ which is obtained by changing the square of, respectively, $1,\,2,\,4$ imaginary units from $-1$ to $1$, but otherwise leaving the multiplication table unchanged.
%from $J(U)$ to $H_2(\mathbb{K}^s)$,
%where $\mathbb{K}=\C,\,\H,\,\bbO$, respectively. 
Furthermore,
the bilinear form on $\g_{-1}$, 
associated to the graded involution (\ref{tilde})
still has the
form $(x,\,y)=\tr{(x \circ y)}$.

To sum up, we have in this section given the 3-grading of $\so(p+1,\,q+1)$ and shown that it is generated by the simple root corresponding to the leftmost node in the Dynkin diagram. %(drawn as on page \pageref{dynkin-sida}).
If we add $n-1 \geq 1$ nodes to the left, then this simple root will instead generate a 5-grading of the resulting algebra $\so(p+n,\,q+n)$. 
Theorem \ref{sats1} 
tells us how the two %generalized Jordan 
triple systems, associated to the 5-graded Lie algebra $\so(p+n,\,q+n)$ and its 3-graded subalgebra $\so(p+1,\,q+1)$, respectively, are related to each other. %The conclusion is that the generalized Jordan triple system associated to
It follows that
$\so(p+n,\,q+n)$
% and 
for $(p,\,q)=(1,\,2),\,(1,\,3),\,(1,\,5),\,(1,\,9)$, %respectively,
is the Lie algebra associated to the generalized Jordan triple system
$H_2(\mathbb{K})^{n}$, for $\mathbb{K}=\R,\,\C,\,\H,\,\bbO$, with the triple product 
\begin{align} \label{tripppel}
(x^a y^b z^c)&=
2\delta^{ab} ((z\circ y)\circ x)^c-2\delta^{ab}((z\circ x) \circ y )^c 
+2\delta^{ab}((x \circ y) \circ z)^c\nn\\
 &\quad -\delta^{ab}(x,\,y)z^c+\delta^{bc}(x,\,y)z^a,
\end{align}
where $a,\,b,\,c=1,\,2,\,\ldots,\,n$ and $(x,\,y)=\tr{(x \circ y)}$. The same holds if we replace these pseudo-orthogonal algebras by the split forms of the corresponding complex Lie algebras, and $\C,\,\H,\,\bbO$ by 
$\C^s,\,\H^s,\,\bbO^s$.

%\newpage

\section{Conclusions} \label{sista}

In this paper we have shown that any Jordan algebra $H_2(\mathbb{K})$ or $H_3(\mathbb{K})$, via (generalized) Jordan triple systems, leads to an infinite sequence of Kac-Moody algebras, labeled by a positive integer $n$. The generalized Jordan triple product is given by (\ref{tripppel}) where $a,\,b,\,c=1,\,2,\,\ldots,\,n$, and $(x,\,y)$ is the bilinear form associated to the graded involution on the Lie algebra in the $n=1$ case.
%generalized Jordan triple system generalizes to an infinite sequence of such triple systems, labeled by a positive integer $n$. 
We have shown that it is given by $(x,\,y)=\tr{(x \circ y)}$ in the $H_2(\mathbb{K})$ case, but not checked it for $H_3(\mathbb{K})$.
However, the bilinear form is well defined by $(e_{\mu},\,\tau(f_{\nu}))=\delta_{\mu \nu}$, and possible to determine if one would like to study the $H_3(\mathbb{K})$ case in detail, as we have done for
$H_2(\mathbb{K})$. This would be an interesting subject of future research.

An important difference between the $H_2(\mathbb{K})$ and
$H_3(\mathbb{K})$ cases is that the Lie algebra associated to $H_2(\mathbb{K})^n$ is 3-graded for $n=1$ and then 5-graded for all $n \geq 2$, while the Lie algebra associated to $H_3(\mathbb{K})^n$ is 3-graded for $n=1$ 
but 7-graded for $n = 2$,
and for $n=3,\,4,\,5,\ldots$,
we get infinitely many subspaces in the grading, since these Lie algebras are infinite-dimensional.
In the affine case, we only get the corresponding current algebra directly in this construction, which means that the central element and the derivation must be added by hand. 
%It would be interesting to find an interpretation of these elements in the Jordan algebra approach. 
The construction might be related to the 'affinization' of generalized Jordan triple systems used in \cite{1988PhLB..209..498G,1989JMP....30..937G} (see also \cite{Goddard:1987ak}).
Finally, concerning the hyperbolic case and further extensions, we hope that our new construction can give more information about these indefinite Kac-Moody algebras, for example $\frake_{10}$ and $\frake_{11}$, which both (but in different approaches) are conjectured to be symmetries underlying M-theory \cite{Damour:2002cu,west:2001as}. In spite of a great interest from both mathematicians and physicists, these algebras are not yet fully understood.

\vspace{1cm}

\noindent
\textbf{Acknowledgments:} I would like to thank Martin Cederwall, Axel Kleinschmidt, Daniel Mondoc, Hermann Nicolai, Bengt E.W. Nilsson, Daniel Persson, Christoffer Petersson and Hidehiko Shimada for discussions.

\appendix

\section{The Lie algebra associated to a generalized \\Jordan triple system}

\label{Associated}

In the end of Section \ref{gradedlieagebras} we saw that any graded Lie algebra with a graded involution gives rise to a generalized Jordan triple system. In this section, we will show the converse, that any generalized Jordan triple system gives rise to a graded Lie algebra with a graded involution. The associated Lie algebra has been defined in different (but equivalent) way by Kantor \cite{Kantor3} and called the \textit{Kantor algebra} \cite{Asano2}.

We recall from Section \ref{gradedlieagebras}
that a generalized Jordan triple system is 
a triple system that satisfies the identity
\begin{align}
(uv(xyz))-(xy(uvz))=((uvx)yz)-(x(vuy)z). \label{GJTS-identity}
\end{align}
For any pair of elements $x,\,y$ in a generalized Jordan triple system $T$, we define the linear map 
\begin{align}
s_{xy}: T \to T,\quad s_{xy}(z)=(xyz).
\end{align}
Thus (\ref{GJTS-identity}) (for all $z$) can be written
\begin{align}
[s_{uv},\,s_{xy}]=s_{(uvx)y}-s_{x(vuy)}.
\end{align}
For any $x \in T$, we also define the linear map 
\begin{align}
v_x : T \to \End{T},\quad v_x(y)=s_{xy},
\end{align}
which we will use in the following subsection.

\subsection{Construction}

Let $T$ be a vector space and set $\tilde{U}_0=\End{T}$. For $k < 0$, define 
$\tilde{U}{}_k$ recursively as the vector space of all linear maps from $T$ to $\tilde{U}{}_{k+1}$. Let $\tilde{U}{}_-$ be the direct sum of all these vector spaces,
\begin{align}
\tilde{U}{}_-= \tilde{U}{}_{-1} \oplus \tilde{U}{}_{-2} \oplus \cdots,
\end{align}
and define a graded Lie algebra structure on $\tilde{U}{}_-$
recursively by the relations
\begin{align}
[u,\,v]&=({\text{ad }u}) \circ v - ({\text{ad }v}) \circ u.
\end{align}
Assume now that $T$ is a generalized Jordan triple system. Let $U_0$ be the subspace of $\tilde{U}_0$ spanned by $s_{uv}$ for all $u,\,v \in T$, and let $U_-$ be the subspace of $\tilde{U}_-$ generated by $v_{x}$ for all $x \in T$.
Furthermore, let $U_+$ be a Lie algebra isomorphic to $U_-$, with the isomorphism denoted by
\begin{align}
\ast : U_- \to U_+, \quad u \mapsto u^{\ast}.
\end{align}
Thus $U_+$ is generated by $v_x{}^\ast$ for all $x \in T$.
Consider the vector space 
\begin{align}
L(T)=U_- \oplus U_0 \oplus U_+. 
\end{align}
We can extend the Lie algebra structures on each of these subspaces to a Lie algebra structure on the whole of $L(T)$, by the relations
\begin{align}
[s_{xy},\,v_z]=v_{(xyz)},\qquad [v_x,\,v_y{}^{\ast}]=s_{xy},
\qquad[s_{xy},\,v_z{}^\ast]=-v_{(yxz)}{}^\ast.
\end{align}
Furthermore, we can
extend the isomorphism $\ast$ between the subalgebras $U_-$ and $U_+$ to a graded involution on the Lie algebra $L(T)$. On $U_+$, it is given by the inverse of the original isomorphism, $(u^{\ast})^{\ast}=u$, 
and on $U_0$ by $s_{xy}{}^{\ast}=-s_{yx}$.
\begin{hej} \label{sats2}
Let $\g$ be
a graded simple Lie algebra, generated by its subspaces $\g_{\pm 1}$, with 
a graded involution $\tau$.
Let $\g_{-1}$ be the generalized Jordan triple system derived from 
$\g$ by 
\begin{align}
(uvw)= [[u,\,\tau(v)],\,w]. 
\end{align}
Then the Lie algebra $L(\g_{-1})$ is isomorphic to $\g$. 
\end{hej}
\Pf 
Define the linear map $\varphi:\g \to L(\g_{-1})$, with $\g_k \to U_k$ for all integers $k$, recursively by
\begin{align} \label{tjofaderittan}
u \in \g_{-} &: 
\quad \varphi(u)(x)=\varphi([u,\,\tau(x)]),\nn\\
s \in \g_0 &:
\quad  \varphi(s)(x)=[s,\,x],\nn\\
\tau(u) \in \g_{+} &: 
\quad  \varphi(\tau(u)) =\varphi(u)^{\ast},
\end{align}
where $x \in \g_{-1}$.
We will show that $\varphi$ is an isomorphism.

\begin{itemize}
\item $\varphi$ \textit{is injective}
\end{itemize}
Suppose that $r$ and $s$ are elements in $\g_{0}$ such that $\varphi(r)=\varphi(s)$. Then $[r-s,\,x]=0$ for all $x \in \g_{-1}$, which means that
$[r-s,\,\g_-]=0$ since $\g_-$ is generated by $\g_{-1}$. But then the proper subspace
\begin{align}
\sum_{k \in \mathbb{N}} (\text{ad }(\g_0 + \g_+))^k (r-s) \subset \g_+ + \g_0
\end{align}
of $\g$ is an ideal.
Since $\g$ is simple, it must be zero, but $r-s$ is an element of this subspace, so $r=s$.
Suppose now that $u$ and $v$ are elements in $\g_{-}$ with $\varphi(u)=\varphi(v)$. 
Then $\varphi([u,\,\tau(x)])=\varphi([v,\,\tau(x)])$ for all $x \in \g_{-1}$, and by induction we can show that this implies $[u-v,\,\tau(x)]=0$ for all $x \in \g_{-1}$. Now we can use the same argument as before (but with $\g_{+}$ replaced by $\g_-$) to show that $u$ and $v$ must be equal. The case $u,\,v \in \g_+$ then easily follows by
\begin{align}
\varphi(\tau(u))-\varphi(\tau(v))=\varphi(u)^\ast - \varphi(v)^\ast
=(\varphi(u)-\varphi(v))^\ast.
\end{align}
\begin{itemize}
\item $\varphi$ \textit{is a homomorphism}
\end{itemize}
It is sufficient to show this when $u \in \g_i$ and $v \in \g_j$ for all integers $i,\,j$ and we will do it by induction over $|i|+|j|$.
One easily checks that
$\varphi([u,\,v])=[\varphi(u),\,\varphi(v)]$ when $|i|+|j| \leq 1$.
Thus suppose that
this is true if $|i|+|j| = p$ for some integer $p \geq 1$. 
For $i,\,j < 0$ we now have
\begin{align}
[\varphi(u),\,\varphi(v)](x)&=
[\varphi(u),\,\varphi(v)(x)]
-[\varphi(v),\,\varphi(u)(x)]\nn\\
&=[\varphi(u),\,\varphi([v,\,\tau(x)])]
-[\varphi(v),\,\varphi([u,\,\tau(x)])]\nn\\
&=\varphi([u,\,[v,\,\tau(x)]]
-[v,\,[u,\,\tau(x)]])=\varphi([u,\,v])(x)
\end{align}
by the assumption of induction in the third step, and by the Jacobi identity in the last one. We use this for the case $i,\,j > 0$, where we have
\begin{align}
\varphi([\tau(u),\,\tau(v)])&=\varphi(\tau([u,\,v]))=\varphi([u,\,v])^\ast=[\varphi(u),\,\varphi(v)]^\ast\nn\\
&= [\varphi(u)^\ast,\,\varphi(v)^\ast]=[\varphi(\tau(u)),\,\varphi(\tau(v))].
\end{align}
Finally, we consider the case where $i \geq 0$ and $j \leq 0$. Again, we show it by induction over $|i|+|j|$, which means $i-j$ in this case. 
One easily checks that it is true when $i=1$ and $j=-1$, so we can assume that $j \leq -2$ (or, analogously, $i \geq 2$).
Then $v$ can be written as a sum of elements $[x,\,y]$ where $x \in \g_{m}$ and $y \in \g_{n}$ for $j < m,\,n < 0$. We consider one such term and, using what we have already proven, we get
\begin{align}
[\varphi(u),\,\varphi([x,\,y])]&=[\varphi(u),\,[\varphi(x),\,\varphi(y)]]\nn\\
&=[[\varphi(u),\,\varphi(x)],\,\varphi(y)]
-[[\varphi(u),\,\varphi(y)],\,\varphi(x)]\nn\\
&=[\varphi([u,\,x]),\,\varphi(y)]
-[\varphi([u,\,y]),\,\varphi(x)]\nn\\
&=\varphi([[u,\,x],\,y]-[[u,\,y],\,x])=\varphi([u,\,[x,\,y]])
\end{align}
by the assumption of induction in the third and fourth steps. 
\begin{itemize}
\item $\varphi$ \textit{is surjective}
\end{itemize}
Since $\varphi$ is a homomorphism, this follows from the fact that $\g$ and $L(\g_{-1})$ are generated by $\g_{\pm 1}$ and $U_{\pm 1}$, respectively.
The proof is complete.
\qed

As we have seen, the theorem is useful when a generalized Jordan triple system 
$T$ happens to be isomorphic to $\g_{-1}$, because it then tells us how to construct $\g$ from $T$.

\subsection{Realization}

The construction of the Lie algebra in the previous subsection may seem rather abstract, where the elements are linear operators acting on vector spaces of other linear operators, which in turn act on other vector spaces, and so on. However, once the Lie algebra is constructed, it can also be realized in a way such that the elements act on the same vector space, but in general non-linearly, and there is a very simple formula for this, as we will see in this subsection.

Let $V$ be the direct sum of (infinitely many) vector spaces 
$V_1,\,V_2,\,\ldots$.
We write an element $v \in V$ as $v = v_1+ v_2 + \cdots$, where $v_k \in V_k,$ for $k=1,\,2,\,\ldots$.
With an operator on $V$ of grade $p$ we mean a map $f: V \to V$ such that for any $i=1,\,2,\,\ldots$, there is a symmetric
$(p_1 + p_2 + \cdots)$-linear map
\begin{align}
F_i : (V_1)^{p_1} \times (V_2)^{p_2} \times \cdots \to V_i,
\end{align}
where $p_1+ 2p_2 + 3p_3 + \cdots = i+p$,
that satisfies 
\begin{align} 
f(v)_i = F_i(v_1,\,v_1,\,\ldots,\,v_1;\,v_2,\,v_2,\,\ldots,\,v_2;\ldots).
\end{align}
We define the composition $f \circ g$ of such an operator $f$ and another operator 
$g$, of grade $q$, as the operator of grade $p+q$ given by 
\begin{align}
(f \circ g)_i (v) &= p_1F_i(g(v)_1,\,v_1,\,\ldots,\,v_1;\,v_2,\,v_2,\,\ldots,\,v_2;\ldots)\nn\\&\quad +p_2F_i(v_1,\,v_1,\,\ldots,\,v_1;\,g(v)_2,\,v_2,\,\ldots,\,v_2;\ldots)
+\cdots
\end{align}
for all $i=1,\,2,\,\ldots$, and a Lie bracket as usual by 
$[f,\,g]=f \circ g - g \circ f$. 
Let $M_p(V)$ be the vector space of all operators on $V$ of grade $p$, and let $M(V)$ be the direct sum of all $M_p(V)$ for all integers $p$ (note that they can also be negative). It follows that $M(V)$ is a graded Lie algebra. It is isomorphic to the Lie algebra of all vector fields $f^i \partial_i$ on $V$, where $f \in M(V)$,
with an isomorphism given by
\begin{align}
f \mapsto -f^i \partial_i. \label{vektorfaltisomomorfi}
\end{align}
Any graded Lie algebra $\g=\g_{-}+\g_0+\g_+$ is isomorphic to a subalgebra of $M(\g_-)$. It can be shown \cite{Kantor3.5,Kantor5} that an injective homomorphism $\chi: \g \to M(\g_{-})$ is given by
\begin{align} 
\label{kantor-formel}
\chi(u) : x \mapsto \left( \frac{{\text{ad }x}}{1-e^{-{\text{ad }x}}}Pe^{-{\text{ad }x}}\right)(u),
\end{align}
where $P$ is the projection onto $U_-$ along $U_0 + U_+$, and the ratio should be considered as the power series
\begin{align} \label{potens-serie}
\frac{{\text{ad }x}}{1-e^{-\text{ad }x}}=
1+\frac{{\text{ad }x}}{2}+\frac{{(\text{ad }x)}^2}{12}-\frac{{(\text{ad }x)}^4}{720}+\cdots.
\end{align}

\subsection{Examples}

We will now illustrate the ideas in two cases, where the generalized Jordan triple system satisfies further conditions.

First of all, we assume that the triple systems are such that if $(xyz)=0$ for all $y,\,z$, then $x=0$. This allows us to identify $x$ with $v_x$ for all $x$ in the triple system, that is, we can identify $U_{-1}$ with $T$, and we can consider any element $[v_x,\,v_y] \in U_{-2}$ as a linear map on $T$, which we denote by $\la x,\,y \ra$.
Since
\begin{align}
[v_x,\,v_y](z)&=[v_x,\,v_y(z)]-[v_y,\,v_x(z)]\nn\\
&=[v_x,\,s_{yz}]-[v_y,\,s_{xz}]=
v_{(xzy)}-v_{(yzx)},
\end{align}
this linear map is given by
\begin{align} \label{mappen}
\la x,\,y \ra (z)=(xzy)-(yzx).
\end{align}
A generalized Jordan triple system is generalized in the sense that this linear map does not have to be zero $-$ in a \textit{Jordan triple system} 
\cite{Jacobson1}, the triple product $(xyz)$ is by definition symmetric in $x$ and $z$.
Accordingly, the Lie algebra associated to a Jordan triple system is 3-graded, 
$\g=\g_{-1}+\g_0+\g_1$, and it can be realized on its subspace $\g_{-1}$ by applying the formula (\ref{kantor-formel}).
Everything that is left from the power series expansion (\ref{potens-serie}) is then the identity map,
\begin{align} 
\frac{{\text{ad }x}}{1-e^{-\text{ad }x}}=1,
\end{align}
and we get
\begin{align} \label{konform_realisering}
u \in \g_{-1} :\quad x &\mapsto Pu = u\nn\\
[u,\,\tau(v)]\in \g_{0} :\quad x &\mapsto P([u,\,\tau(v)] - [x,\,[u,\,\tau(v)]])=-[x,\,[\tau(u),\,v]]\nn\\
&\quad=[s_{uv},\,x]=(uvx),\nn\\
\tau(u) \in \g_{1} :\quad x &\mapsto P(\tau(u) - [x,\,\tau(u)] + \tfrac{1}{2}[x,\,[x,\,\tau(u)]])=\tfrac{1}{2}[x,\,[x,\,\tau(u)]]\nn\\
&\quad=-\tfrac{1}{2}[s_{xu},\,x]=-\tfrac{1}{2}(xux).
\end{align}
This is the \textit{conformal realization} of $\g$ on $\g_{-1}$.

We now turn to \textit{Kantor triple systems} \cite{Allison} (or generalized Jordan triple systems of \textit{second order} \cite{Kantor3}). These are generalized Jordan triple systems that in addition to the condition (\ref{GJTS-identity}) satisfy the identity
\begin{align}
\label{ktsdef2}
\langle \langle u,\,v \rangle (x),\,y \rangle = 
\langle (yxu),\,v \rangle - \langle (yxv),\,u \rangle.
\end{align}
It follows that the Lie algebra associated to a Kantor triple system is 5-graded, and the only part of (\ref{potens-serie}) that we have to keep is
\begin{align} 
\frac{{\text{ad }x}}{1-e^{-\text{ad }x}}=
1+\frac{{\text{ad }x}}{2}.
\end{align}
Then we get
\begin{align} \label{kantoroperatorer}
[u,\,v]\in \g_{-2}:\quad z + Z&\mapsto \langle u,\, v \rangle,\nn\\
u\in \g_{-1}:\quad z + Z &\mapsto u + \tfrac{1}{2} \langle u,\, z \rangle,\nn\\
[u,\,\tau(v)]\in \g_{0}:\quad z + Z&\mapsto (uvz) 
-\langle u,\, Z (v) \rangle,\nn\\
\tau({u})\in \g_{1}:\quad z + Z &\mapsto 
-\tfrac{1}{2}(zuz)- Z (u)\nn\\
                &\quad +\tfrac{1}{12}\langle (zuz),\,z \rangle-\tfrac{1}{2}\langle Z (u),\,z \rangle,\nn\\
\tau({u}),\,\tau(v)]\in \g_{2}:\quad z+Z &\mapsto -\tfrac{1}{6}(z\langle u,\,v\rangle(z)z)
-Z(\langle u,\,v \rangle (z))\nn\\
                &\quad + \tfrac{1}{24} \langle (z\langle u,\,v\rangle(z)z),\, z \rangle
+\langle Z(u),\,Z(v) \rangle,
\end{align}
where $z \in \g_{-1}$ and $Z \in \g_{-2}$,
which is the same realization as in \cite{Palmkvist:2005gc}, apart from a rescaling of the elements in $\g_{-2}$ by a factor of two.

\label{associated}

\bibliographystyle{utphysmod2}

%\bibliography{biblio}

\begin{thebibliography}{10}

\bibitem{Jordan}
P.~Jordan,  {\em Uber eine {K}lasse nichtassociativer hyperkomplexer
  {A}lgebren}, Nachr. Ges. Wiss. G\"ottingen 569--575 (1932).

\bibitem{Jordneuwig}
P.~Jordan, J.~von Neumann and E.~Wigner,  {\em On an algebraic generalization
  of the quantum mechanical formalism}, Ann. Math. {\bf 35}, 29--64 (1934).

\bibitem{Gunaydin:1975mp}
M.~G{\"u}naydin,  {\em Exceptional realizations of {L}orentz group:
  Supersymmetries and leptons}, Nuovo Cim. {\bf A29}, 467
(1975).
%%CITATION = NUCIA,A29,467;%%.

\bibitem{Gunaydin7}
M.~G{\"u}naydin, {\em in {E}lementary Particles and the Universe: Essays in the
  Honor of {G}ell-{M}ann, ed. by {J}.{H}. {S}chwarz}.
\newblock Cambridge, 1991.

\bibitem{Kantor1}
I.~L. Kantor,  {\em Classification of irreducible transitively differential
  groups}, Soviet Math. Dokl. {\bf 5}, 1404--1407 (1964).

\bibitem{Koecher}
M.~Koecher,  {\em Imbedding of {J}ordan algebras into {L}ie algebras {I}},
  Amer. J. Math. {\bf 89}, 787--816 (1967).

\bibitem{Tits1}
J.~Tits,  {\em Une classe d'alg{\`e}bres de {L}ie en relation avec les
  alg{\`e}bres de {J}ordan}, Indag. Math. {\bf 24}, 530--534 (1962).

\bibitem{Kantor3.5}
I.~L. Kantor,  {\em Some generalizations of {J}ordan algebras}, Trudy Sem.
  Vect. Tens. Anal. {\bf 16}, 407--499 (1972).

\bibitem{Tits}
J.~Tits,  {\em Alg{\`e}bres alternatives, alg{\`e}bres de {J}ordan et
  alg{\`e}bres de {L}ie exceptionelles}, Indag. Math. {\bf 28}, 223--237
  (1966).

\bibitem{Vinberg}
E.~B. Vinberg,  {\em A construction of exceptional simple {L}ie groups}, Tr.
  Semin. Vektorn. Tensorn. Anal. {\bf 13}, 7--9 (1966).

\bibitem{Kantorartikel}
I.~L. Kantor,  {\em Models of exceptional {L}ie algebras}, Soviet Math. Dokl.
  {\bf 14}, 254--258 (1973).

\bibitem{Sudbery2}
C.~H. Barton and A.~Sudbery,  {\em Magic squares and matrix models of {L}ie
  algebras}, Adv. Math. {\bf 180}, 596--647 (2003)
  [\href{http://www.arXiv.org/abs/math.RA/0203010}{{\tt math.RA/0203010}}].

\bibitem{Gunaydin}
M.~G{\"u}naydin, K.~Koepsell and H.~Nicolai,  {\em Conformal and quasiconformal
  realizations of exceptional {L}ie groups}, Commun. Math. Phys. {\bf 221},
  57--76 (2001)
[\href{http://www.arXiv.org/abs/hep-th/0008063}{{\tt hep-th/0008063}}].
%%CITATION = HEP-TH 0008063;%%.

\bibitem{Palmkvist:2005gc}
J.~Palmkvist,  {\em A realization of the {L}ie algebra associated to a {K}antor
  triple system}, J. Math. Phys. {\bf 47}, 023505 (2006)
[\href{http://www.arXiv.org/abs/math.RA/0504544}{{\tt math.RA/0504544}}].
%%CITATION = MATH/0504544;%%.

\bibitem{Kantor5}
I.~Kantor,  {\em On a vector field formula for the {L}ie algebra of a
  homogeneous space}, Journal of Algebra {\bf 235}, 766--782 (2001).

\bibitem{Kac}
V.~G. Kac, {\em Infinite dimensional {L}ie algebras}.
\newblock 3rd edition, Cambridge University Press, 1990.

\bibitem{Sudbery}
A.~Sudbery,  {\em Division algebras, (pseudo)orthogonal groups and spinors},
  Jour. Phys. A: Math. Gen. {\bf 17}, 939--955 (1984).

\bibitem{Mondocavhandling}
D.~Mondoc, {\em Kantor Triple Systems}.
\newblock Doctoral thesis, Lund University, 2002.

\bibitem{Kantor4}
I.~L. Kantor and I.~M. Skopets,  {\em Some results on {F}reudenthal triple
  systems}, Sel. Math. Sov. {\bf 2}, 293--305 (1982).

\bibitem{Faulkner}
J.~R. Faulkner,  {\em A construction of {L}ie algebras from a class of ternary
  algebras}, Trans. Amer. Math. Soc. {\bf 155}, 397--408 (1971).

\bibitem{1988PhLB..209..498G}
N.~{G{\"u}naydin} and S.~J. {Hyun},  {\em {Affine exceptional Jordan algebra
  and vertex operators}}, Physics Letters B {\bf 209}, 498--502 (Aug., 1988).

\bibitem{1989JMP....30..937G}
M.~{G{\"u}naydin},  {\em {Vertex operator construction of nonassociative
  algebras and their affinizations}}, Journal of Mathematical Physics {\bf 30},
  937--942 (Apr., 1989).

\bibitem{Goddard:1987ak}
P.~Goddard, W.~Nahm, D.~I. Olive, H.~Ruegg and A.~Schwimmer,  {\em {F}ermions
  and octonions}, Commun. Math. Phys. {\bf 112}, 385
(1987).
%%CITATION = CMPHA,112,385;%%.

\bibitem{Damour:2002cu}
T.~Damour, M.~Henneaux and H.~Nicolai,  {\em ${E}_{10}$ and a 'small tension
  expansion' of {M}-theory}, Phys. Rev. Lett. {\bf 89}, 221601 (2002)
[\href{http://www.arXiv.org/abs/hep-th/0207267}{{\tt hep-th/0207267}}].
%%CITATION = HEP-TH 0207267;%%.

\bibitem{west:2001as}
P.~C. West,  {\em $E_{11}$ and {M}-theory}, Class. Quant. Grav. {\bf 18},
  4443--4460 (2001)
[\href{http://www.arXiv.org/abs/hep-th/0104081}{{\tt hep-th/0104081}}].
%%CITATION = HEP-TH/0104081;%%.

\bibitem{Kantor3}
I.~L. Kantor,  {\em Formulas for infinitesimal operators of the {L}ie algebra
  of a homogeneous space}, Trudy Sem. Vect. Tens. Anal. {\bf 17}, 243--249
  (1974).

\bibitem{Asano2}
H.~Asano and S.~Kaneyuki,  {\em Graded {L}ie algebras and generalized {J}ordan
  triple systems}, Nagoya Math. J. {\bf 112}, 81--115 (1988).

\bibitem{Jacobson1}
N.~Jacobson,  {\em Lie and {J}ordan triple systems}, Amer. J. Math. {\bf 71},
  149--170 (1949).

\bibitem{Allison}
B.~N. Allison and J.~R. Faulkner,  {\em Elementary groups and invertibility for
  {K}antor pairs}, Comm. Algebra {\bf 27}, 519--556 (1999).

\end{thebibliography}

\providecommand{\href}[2]{#2}\begingroup\raggedright\endgroup

\end{document}